\documentclass[sigplan,10pt,nonacm]{acmart}
\renewcommand\footnotetextcopyrightpermission[1]{}

\acmSubmissionID{1595}

\usepackage{tikz}
\usepackage{amsmath}

\usepackage{filecontents}

\usepackage{enumitem}
\setlist[itemize]{leftmargin=1.2em}

\usepackage{xcolor}    
\usepackage{booktabs}
\usepackage{amsthm}
\usepackage{algorithm}
\usepackage[noend]{algpseudocode}
\newcommand{\systemname}{\textsc{ProServe}}
\newcommand{\localsched}{SlideBatching}
\newcommand{\globalsched}{GoRouting}
\newcommand{\on}{\textcolor{green}{\Large\textbf{\checkmark}}}
\newcommand{\off}{\textcolor{red}{\LARGE\textbf{\texttimes}}} 

\begin{document}

\title{\systemname{}: Unified Multi-Priority Request Scheduling for LLM Serving}

\author{
{\rm Weizhe Huang}$^1$, 
{\rm Tao Peng}$^1$, 
{\rm Tongxuan Liu}$^1$,
{\rm Donghe Jin}$^1$,
{\rm Meng Kang}$^1$,
{\rm Xianzhe Dong}$^2$,
{\rm Ke Zhang}$^3$
\\
$^1$JD.com \quad $^2$USTC \quad $^3$Unaffiliated
} 

\renewcommand{\shortauthors}{Weizhe Huang et al.}

\begin{abstract}

The widespread deployment of large language models (LLMs) for interactive applications necessitates serving systems that can handle thousands of concurrent requests with diverse Service Level Objective (SLO) requirements. A critical yet often overlooked dimension in this context is the \textit{inherent priority difference among clients}; for instance, business-critical functions demand higher performance guarantees, as fulfilling such requests yields significantly greater business value. However, existing LLM serving schedulers fail to jointly optimize for both SLO attainment and client-level priorities.

To bridge this gap, we first \textit{formalize multi-priority request scheduling as a service gain maximization problem}, where satisfying latency requirements for requests of different priorities contributes varying gain. We propose \systemname{}, a unified two-tier scheduling framework designed to maximize overall service gain. 
At the engine layer, \localsched{} dynamically adapts batch formation under varying loads, employing a sliding boundary mechanism to balance latency and priority differentiation. Considering potential preemption, efficient block management adopts asynchronous offloading, pipelined reloading, and adaptive copy-budget control to overlap computation with host-device block transfers.
At the service layer, \globalsched{} performs gain-oriented and capability-aware dispatching across distributed instances, proactively reserving capacity for future high-priority or long requests.
Extensive evaluation on four open-source and one industrial dataset shows that \systemname{} outperforms state-of-the-art baselines, improving system gain by up to 35\% and SLO attainment by up to 52\%.

\end{abstract}

\maketitle

\section{Introduction}

Large language models (LLMs)~\cite{deepseek, qwen3, llama} have become foundational to a wide range of interactive applications, from chatbots~\cite{chatgpt} to autonomous agents~\cite{karim2025transforming}. As these LLM services are deployed at scale, they need to handle thousands of concurrent online requests with stringent and heterogeneous Service Level Objective (SLO) requirements~\cite{tang2025scorpio, distserve}. 

\begin{figure}[t]
  \centering
  \includegraphics[width=\linewidth]{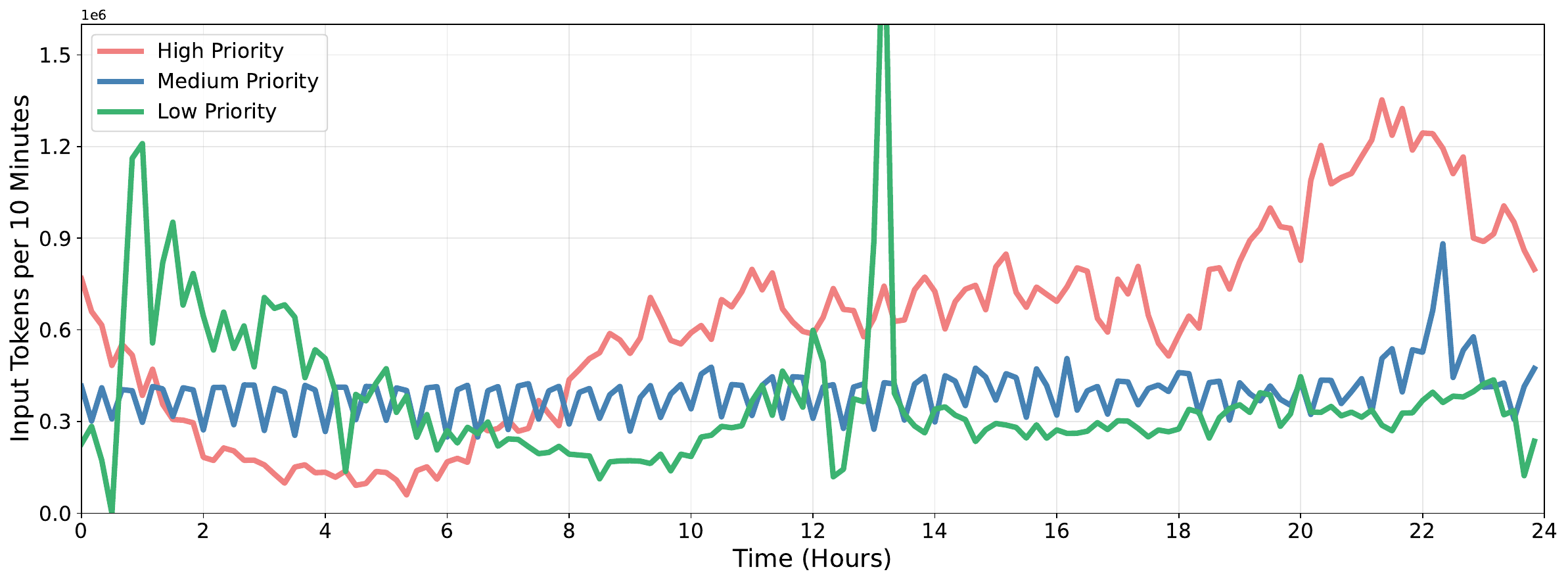}
  \caption{Workload trace of online requests with different priorities from our real-world industrial dataset.}
  \label{fig:multi_priority_load}
\end{figure}

Beyond SLO diversity, a critical yet often neglected dimension is the \textbf{inherent priority difference among the clients themselves}. In real‑world enterprise scenarios, for instance, business‑critical functions demand higher performance guarantees than non‑critical ones. To illustrate this, Figure~\ref{fig:multi_priority_load} presents a real workload trace from our industrial dataset, which shows that requests of different priorities exhibit distinct arrival patterns and load dynamics. Moreover, successfully serving a high‑priority request typically yields significantly greater business value than serving a low‑priority one. Therefore, an effective serving system must not only respect per‑request SLOs, but also differentiate service based on client priority.

However, existing LLM serving schedulers fail to jointly account for both SLO diversity and client‑level priorities. One line of work~\cite{tang2025scorpio, hyperflexis, slos_serve} addresses SLO heterogeneity by implicitly prioritizing requests with tighter deadlines, yet overlooks the inherent priority differences among the clients themselves. 
Another line of work~\cite{wang2025echo, bros, sun2025hygen} focuses on co‑scheduling online and offline tasks, treating all online requests as uniformly high‑priority (to be served with SLO guarantees) and offline ones as best‑effort low‑priority tasks without explicit latency requirements. This design makes such approaches inapplicable to our scenario, where both high‑ and low‑priority requests carry their own latency requirements.
Other works~\cite{sun2024llumnix, vtc} consider online client priority but inadequately account for SLO attainment. Llumnix~\cite{sun2024llumnix} merely reserves static memory without providing explicit latency guarantees. Weighted VTC~\cite{vtc}, inspired by Linux's Completely Fair Scheduler (CFS)~\cite{cfs}, enforces token‑based proportional fairness across priority classes but cannot explicitly satisfy per‑request latency targets, which are critical in LLM serving.
Consequently, there is no principled framework that simultaneously accounts for client priority and SLO attainment.


To bridge this gap, we first \textbf{formulate the multi-priority request scheduling problem as a service gain maximization task} (\textsection\ref{sec:priority_objective}), where satisfying a high-priority\footnote{Throughout this paper, request priority refers to the priority of the originating client.} request's latency requirement contributes substantially more service gain than a low-priority one. We then jointly account for request priority and SLO requirements and propose a novel \emph{Token‑level Deadline‑aware Gain (TDG)} function (\textsection\ref{sec:priority_objective}) that quantifies the gain obtained from meeting the SLO of a specific-priority request. This formula explicitly captures the inherent differences of gain across different priorities, while respecting the individual latency target for each request.

However, solving this problem introduces significant challenges. 
First, two fundamental design choices for balancing latency and priority differentiation arise: how to allocate resources across priority classes, and how to schedule requests with different priorities. Regarding resource allocation, naively dedicating separate resources per priority class leads to waste and insufficiency under dynamic loads, motivating our co-location design. Given co-location, a trade-off exists between minimizing overall latency and favoring priority in unified scheduling. Naive policies like First-Come-First-Served (FCFS)~\cite{vllm} or existing SLO-aware policies~\cite{tang2025scorpio, bros, slos_serve} fail to differentiate priorities, while strict priority-first starves low-priority requests. Both result in suboptimal total system gain (\textsection\ref{sec:latency_priority_tradeoff}).
Second, we observe that the effectiveness of different batch scheduling policies varies significantly with load. Under dynamic workloads, static scheduling policies struggle to adapt across load levels, limiting their effectiveness (\textsection\ref{sec:static_scheduler}).
Third, priority-aware preemption increases KV cache pressure and makes eviction more frequent. Host-device transfers must be carefully coordinated with scheduling to avoid moving memory overhead onto the critical path (\textsection\ref{sec:memory_manage}).
Fourth, in distributed deployments, existing global dispatchers (e.g., least‑load) lack awareness of request priority and suffer from the \textit{over‑balancing} issue. This may cause them to fail to accommodate future high‑priority or long requests even when sufficient capacity is potentially available (\textsection\ref{sec:limitation_global}). 

To address these challenges, we present \systemname{} (\textsection\ref{sec:design}), a unified two-tier scheduling framework designed to maximize service gain from multi-priority requests. 
At the engine layer, we introduce \localsched{} (\textsection\ref{sec:local_sched}), a local batch scheduler that dynamically adapts its policy based on real-time load. It adaptively partitions the request queue into urgent and non-urgent subsets and applies tailored strategies. 
Moreover, we implement efficient block management (\textsection\ref{sec:block_manage}) with asynchronous offloading, pipelined reloading, and adaptive copy-budget control to reduce eviction and reload overhead.
At the service layer, we design \globalsched{} (\textsection\ref{sec:global_sched}), a global request router that performs gain-oriented, capability-aware dispatching across distributed instances. It maintains awareness of local scheduler states and proactively reserves capacity for future high-priority or long requests.
We evaluate \systemname{} against multiple common and state‑of‑the‑art schedulers across four open‑source datasets and one large‑scale real‑world industrial dataset. Extensive experiments demonstrate that \systemname{} consistently and significantly outperforms state-of-the-art baselines, improving system gain by up to 35\% and boosting overall SLO attainment by up to 52\%.

Our main contributions are summarized as follows:
\begin{itemize}
    \item We formally define the multi-priority scheduling scenario and formulate it as a service gain maximization problem. To this end, we propose a novel gain function TDG, which quantifies priority-based gain while evaluating token delivery against deadlines.

    \item At the engine layer, we introduce \localsched{}, a local batch scheduler that dynamically adjusts request ordering strategies based on real-time loads.
    We also introduce an efficient block management mechanism to minimize the overhead of host-device block swapping upon eviction.

    \item At the service layer, we design \globalsched{}, a global router that proactively monitors instance and request states and employs a gain-oriented, capability-aware routing policy.

    \item We demonstrate through extensive experiments that \systemname{} achieves superior and robust performance.
\end{itemize}

\section{Characterizing Request Priority and Service Objectives} \label{sec:priority_objective}


\begin{table}[t]
\centering
\resizebox{\linewidth}{!}{
\begin{tabular}{c|c|c|c|c}
\toprule
\textbf{Gain Function}  & \textbf{Weighted SLO} & \textbf{Tempo~\cite{zhang2025tempo}} & \textbf{TA-SLO} & \textbf{TDG}\\
\midrule
Request Priority  & \on & \off & \on & \on \\
\midrule
Per-Token Latency  & \off & \on & \on & \on \\
\midrule
First/Decode Token Import.  & \off & \on & \on & \on \\
\midrule
Robust to Discard/Postpone & \off/\off & \on/\off & \on/\off & \on/\on \\
\bottomrule
\end{tabular}
}
\caption{The feature comparison of different per-request gain functions. ``Import.'' denotes ``Importance''.}
\label{table:comparison_metric}
\end{table}

Existing scheduling methods typically map request attributes (e.g., sequence length~\cite{fastserve, learning_to_rank, TetriInfer, prefillonly}, SLO constraints~\cite{tang2025scorpio, hyperflexis, slos_serve}, online/offline type~\cite{wang2025echo, bros, sun2025hygen}) to execution priorities. However, they often fail to account for the inherent priority differences among the online clients themselves. 
To address this more general scenario, we formalize this \textit{multi-priority request scheduling problem} as follows:

\noindent \textbf{Objective Definition.}
\label{def:main_problem}
\emph{
Let $\mathcal{P}=\{1,\dots,P\}$ denote a finite set of priority levels.
Each request $r$ is assigned a priority class $p(r)\in\mathcal{P}$.
Let $w:\mathcal{P}\rightarrow\mathbb{R}_{>0}$ be a priority-weight mapping, and define $w_{p(r)}:=w(p(r))$ as the priority weight of request $r$ (optionally, $w_1\ge\dots\ge w_P$).
If request $r$ meets its latency target, the system accrues gain $f(r)$.
The objective is to maximize total gain over served requests $R$:
$\max F(R)=\max \sum_{r\in R} f(r)$.
}

The core issue then shifts to defining a proper per‑request gain function $f(r)$ based on $w_{p(r)}$ and its own $\text{SLO}_r$.

\noindent \textbf{Strawman Proposal 1: Weighted SLO Attainment.}
An intuitive idea is to weight the standard SLO attainment by priority:
\begin{equation}
    f_{W}(r) = w_{p(r)} \cdot \mathbb{I}[TTFT_r < TTFT_{SLO}^r, TPOT_r < TPOT_{SLO}^r],
\end{equation}
However, this formulation suffers from three key drawbacks:
(1) \textit{Discard-or-Postpone Trick}: Since gain is awarded only if both TTFT and TPOT SLOs are met, the system can immediately discard or indefinitely postpone any request whose TTFT SLO is deemed unattainable, as the gain for that request is already lost. This trick, while potentially improving the metric, significantly degrades user experience.
(2) \textit{Insensitivity to Per-Token Latency}: As an average metric, TPOT obscures the variability in per-token delivery times. For instance, a request with high initial latency but very fast subsequent tokens can yield the same TPOT as one with uniformly moderate latency. Consequently, both would attain the same gain $f_{W}(r)$, despite offering substantially different user experiences.
(3) \textit{Undifferentiated Importance of First and Decode Token}: In practice, TTFT and TPOT reflect different dimensions of service quality. TTFT measures the system's initial responsiveness, whereas TPOT reflects the output fluency. A single, combined SLO condition fails to account for their differing impacts on the overall user experience.

\noindent \textbf{Refined Proposal 2: Token-level Gain with TBT.}
To overcome these issues, we shift from request‑level SLO attainment to a token‑level gain function that aggregates the timely delivery of each output token. Inspired by prior work~\cite{zhang2025tempo,agrawal2024etalon}, our refined proposal is to replace TPOT with Time Between Tokens (TBT), leading to the initial Token-level Accumulated SLO (TA-SLO) formulation:
\begin{equation}
\begin{split}
    f_{TA-SLO}(r) = w_{p(r)} \cdot (w_{p} \cdot \mathbb{I}[TTFT_r < TTFT_{SLO}^r] + \\
    \sum_i w_{d} \cdot \mathbb{I}[TBT_{r,i}<TBT_{SLO}^r ] ),
\end{split}
\end{equation}
where $TBT_{r,i} = t_{r,i} - t_{r,i-1}$ is the interval between consecutive tokens and $t_{r,i}$ denotes the output time of the $i$-th token of the request $r$. Here, $w_{p}$ and $w_{d}$ weight the importance of the first versus subsequent tokens.
However, \textit{Postponed Decoding Trick} also persists in this definition. If a token is already detected to miss its TBT SLO, the system might intentionally delay its output to make the next token's TBT easier to achieve. This distorts the gain calculation and stems from the negative monotonicity of the TBT metric: completing one token earlier can negatively impact the TBT SLO attainment of the next.

\noindent \textbf{Our Final Proposal: Token-level Deadline-aware Gain (TDG).}
To overcome the above limitations, we make a fundamental shift in perspective: the gain for a token is interpreted as a binary indicator of whether its output timing harms user experience. Specifically, the individual token deadline represents the latest acceptable output time that does not degrade perceived quality. A token delivered after this deadline yields no gain (as it harms user experience), whereas earlier completion does not increase the gain for that token. 
Based on this insight, we replace the varying deadlines adopted by~\cite{zhang2025tempo,agrawal2024etalon} with a fixed deadline and introduce the Token-level Deadline-aware Gain (TDG).

\begin{equation}
\begin{split}
    f_{TDG}(r) &= \sum_i w_r(i) \cdot \mathbb{I}[t_{r,i} < deadline_{r,i}],\\
    deadline_{r,i} &= TTFT_{SLO}^r+(i-1)\cdot TPOT_{SLO}^r, \\
    w_r(i) &= \begin{cases}
 w_{p} \cdot w_{p(r)} & ,\text{ if } i=1 \\
 w_{d} \cdot w_{p(r)} &  ,\text{otherwise}
\end{cases},
\end{split}
\end{equation}
where $w_r(i)$ is the gain for delivering the $i$‑th token before its deadline. Weights $w_p$ and $w_d$ distinguish the importance of the first versus subsequent tokens, scaled by priority weight $w_{p(r)}$. This mapping can be adapted based on application-specific requirements.

A simple comparison of TDG against other gain functions is presented in Table~\ref{table:comparison_metric}, and a more detailed comparison is shown in Appendix~\ref{sec:comparison}.
Furthermore, TDG establishes clear \textit{monotonicity properties}:
(1) \textit{Positive Impact of Early Completion.}
Although a token completed early gains no extra direct benefit, it increases the slack for subsequent tokens.
Potential generation stalls caused by early output can be mitigated by the smoothing buffer mechanism~\cite{wang2024revisiting} that can be seamlessly integrated into the front-end of LLM serving systems. This mechanism automatically caches sequentially output tokens and presents them at any predetermined intervals whenever tokens are available in the buffer\footnote{In this work, we apply buffering~\cite{wang2024revisiting} to all tokens, including the first token. If the first token cannot be buffered, the deadline can be redefined as $deadline_{r,i} = \min\{TTFT_r, TTFT_{SLO}^r\} + (i-1)\cdot TPOT_{SLO}^r$, which can also align with aforementioned features.}. 
(2) \textit{Negative Impact of Late Completion.}
Because deadlines are fixed and independent, a late token directly reduces the available time for subsequent ones, creating a risk of deadline‑miss propagation. This prevents the \textit{Infinite‑Postpone Trick} and discourages request discarding, as doing so forfeits all potential future gain.

\noindent \textbf{Optimization Hardness.}
Maximizing total TDG is NP-hard even in a restricted single-instance setting. Specifically, by disabling decode-side constraints and batching, the problem subsumes \(1\mid r_j,d_j\mid\sum_j w_jU_j\) as a special case. Therefore, the full multi-instance optimization is at least as hard, which motivates our heuristic scheduler design. A formal proof is provided in Appendix~\ref{sec:appendix_complexity}.

\section{Motivation and Challenges}

\subsection{Balancing Latency and Priority Differentiation} \label{sec:latency_priority_tradeoff}

\begin{figure}[t]
  \centering
  \includegraphics[width=\linewidth]{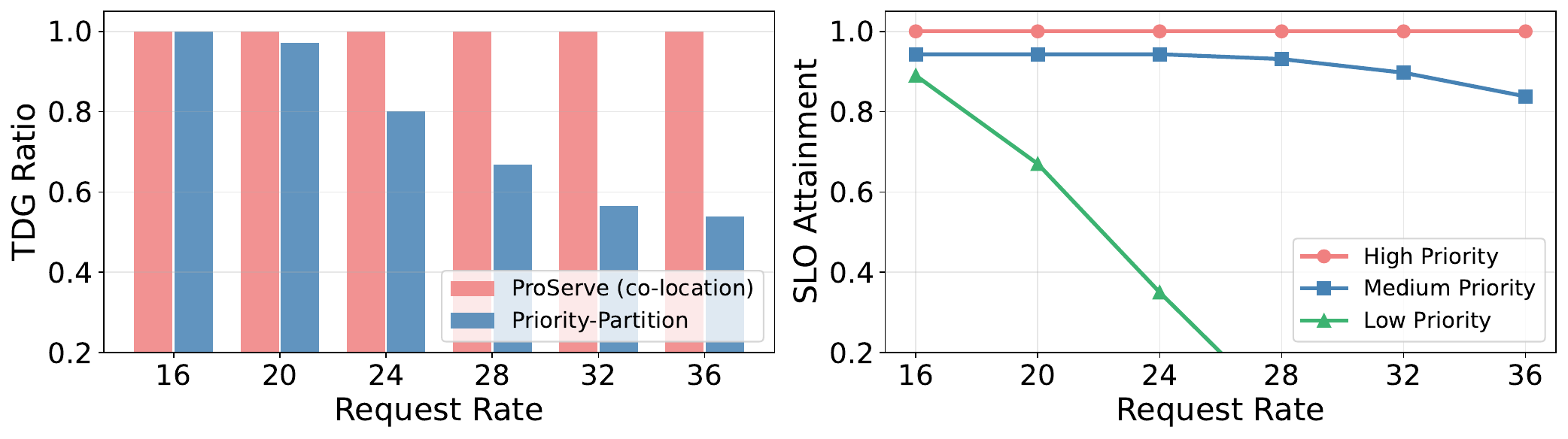}
  \caption{Left: Performance of priority disaggregation versus priority co-location on our industrial dataset. Right: Performance of different priorities under disaggregation.}
  \label{fig:priority_disagg_vs_agg}
\end{figure}

\begin{figure}[t]
  \centering
  \includegraphics[width=\linewidth]{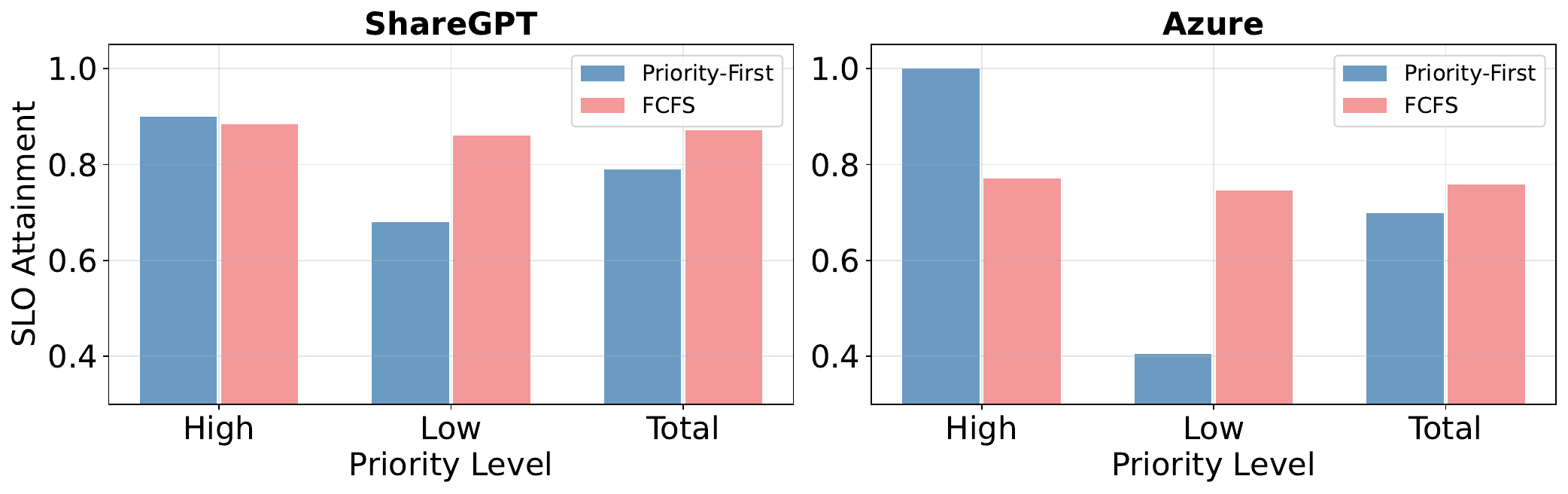}
  \caption{Performance of requests with different priorities in strict Priority-First scheduling and FCFS-based Sarathi~\cite{sarathi-serve}.}
  \label{fig:priority_vs_fcfs}
\end{figure}

\noindent\textbf{Resource Allocation: Partition vs. Co-location.}
A naive approach to serving multi-priority requests is dedicating separate resources per priority class. However, as shown in Figure~\ref{fig:priority_disagg_vs_agg} (left), on our industrial dataset, statically partitioning resources based on the average load of each priority class yields much lower total gain than our co-location design \systemname{}. The right panel further shows the SLO attainment of each priority class under static partitioning during a specific time window. While high-priority requests achieve good SLO attainment, medium- and low-priority ones degrade significantly. The root cause lies in the dynamic load variations across priority classes (Figure~\ref{fig:multi_priority_load}). Static partitioning can lead to resource waste or insufficiency (e.g., low-priority clusters overload, whereas high-priority ones remain idle). Furthermore, dynamically adjusting resources at runtime risks instance downtime and instability. These limitations motivate our exploration of co-locating requests of different priorities to maximize cluster resource utilization.

\noindent\textbf{Scheduling Policy: Priority-First vs. FCFS.}
Our objective gain function incorporates latency-related terms and the priority weights, indicating that both factors must be considered in scheduling. However, an inherent trade-off exists between minimizing overall latency and strictly favoring high-priority requests.
As shown in Figure~\ref{fig:priority_vs_fcfs}, a strict priority-first policy always prioritizes high-priority requests. While this approach significantly improves SLO attainment for high-priority requests, the severe imbalance in computational resource allocation leads to poor overall latency guarantees for the system. Conversely, Sarathi~\cite{sarathi-serve}, a mainstream FCFS-based policy, can achieve overall SLO attainment but fails to provide differentiated service quality across priority classes. In practice, however, the latency of high-priority requests matters more for total system gain.
Thus, co-location further raises a key challenge: how to design an effective scheduler that jointly balances overall latency and priority differentiation to maximize total gain.

\subsection{The Adaptive Deficit of Static Schedulers Under Dynamic Workloads} \label{sec:static_scheduler}

For scheduler design, each iteration must make two key decisions: (1) \textit{request admission order} and (2) \textit{batch capacity}.
For the first, existing studies predominantly adopt fixed scheduling policies, such as FCFS~\cite{vllm}, earliest-deadline-first (EDF)~\cite{bros, fairbatching}, and shortest-job-first (SJF)~\cite{fastserve, learning_to_rank}, as well as their variants. For the second, existing inference engines such as vLLM~\cite{vllm} and xLLM~\cite{liu2025xllm} typically predefine a static token budget (e.g., \texttt{max\_num\_batched\_tokens} in vLLM) or sequence limit (e.g., \texttt{max\_num\_seqs} in vLLM) and keep it unchanged throughout scheduling.
However, online workloads are volatile and unpredictable, with both request intensity and priority mix varying over time (Figure~\ref{fig:multi_priority_load}).
We show that static ordering policies often lack sufficient adaptability, and their effectiveness is sensitive to chosen batch capacity.

\begin{figure}[t]
  \centering
  \includegraphics[width=\linewidth]{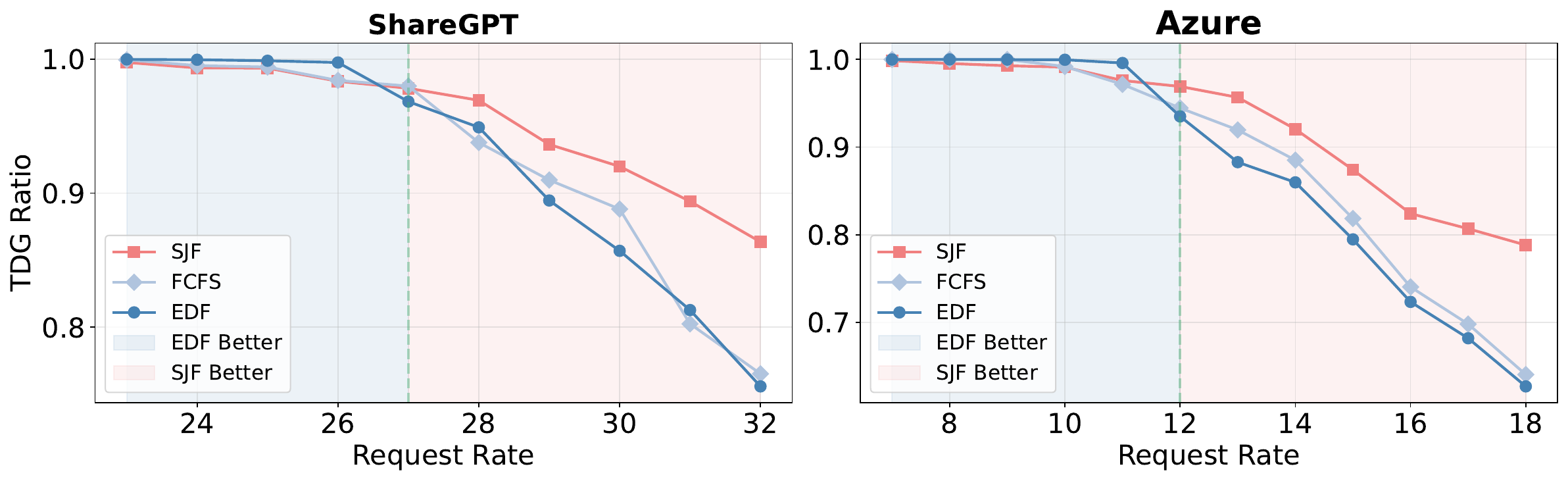}
  \caption{Performance of different request sorting strategies under the fixed token budget.}
  \label{fig:deadline_vs_sjf_same_token_budget}
\end{figure}

\begin{figure}[t]
  \centering
  \includegraphics[width=\linewidth]{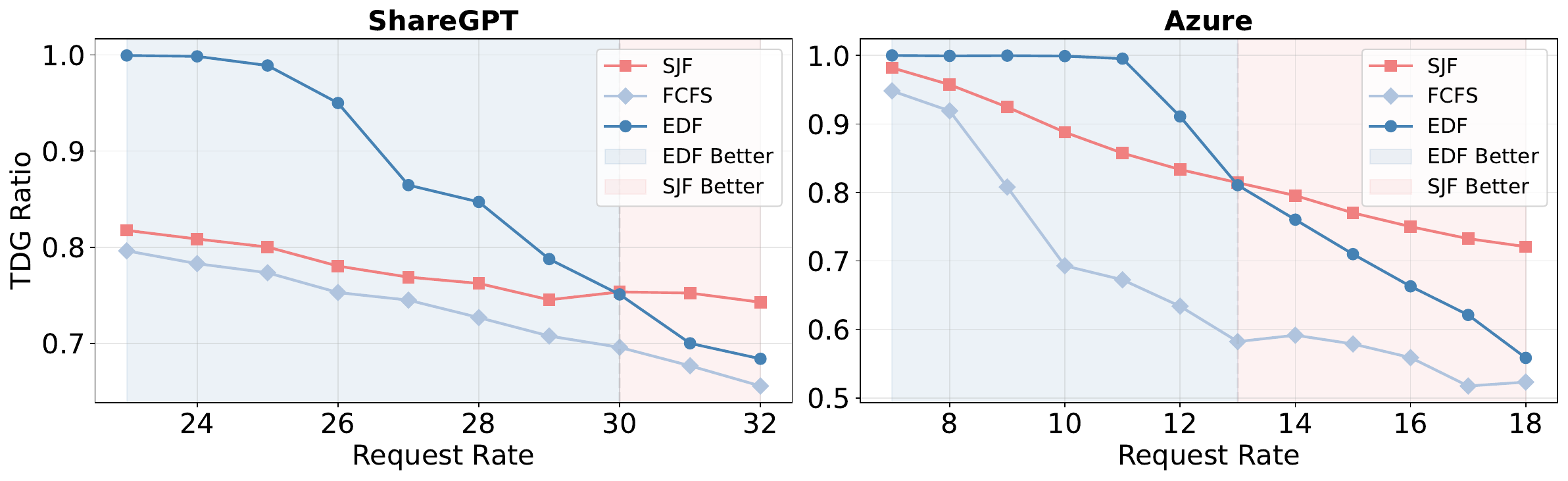}
  \caption{Performance of different request sorting strategies under the fixed batch size.}
  \label{fig:deadline_vs_sjf_same_batch_size}
\end{figure}

\begin{figure}[t]
  \centering
  \includegraphics[width=\linewidth]{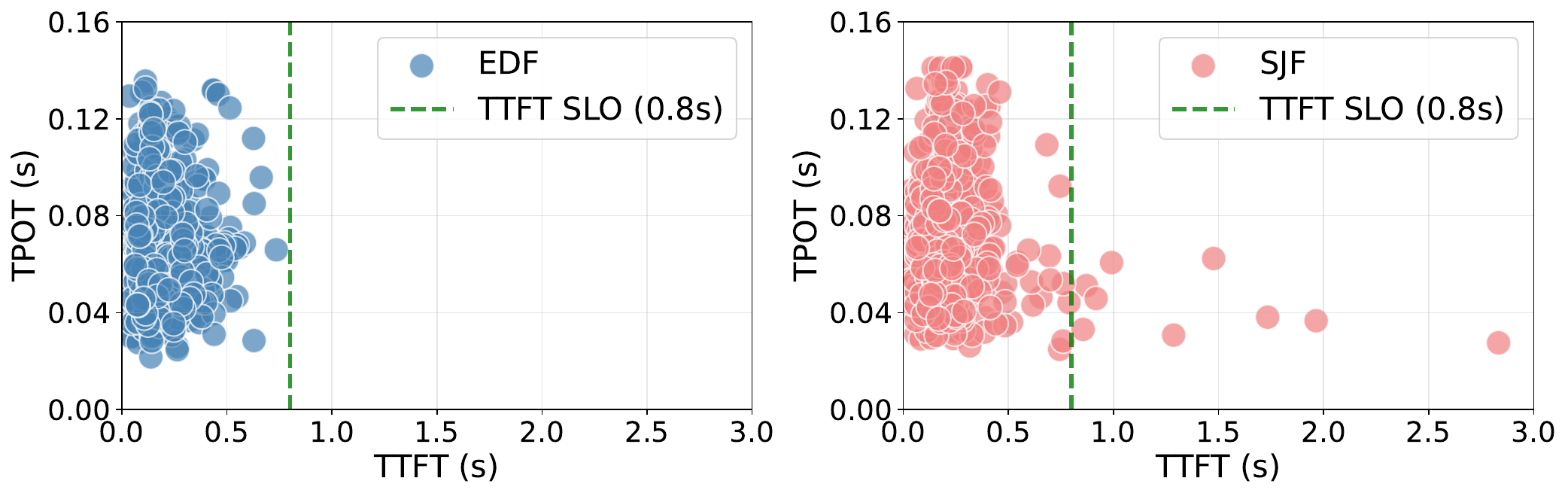}
  \caption{Distributions of TTFT and TPOT under different scheduling policies in low-load scenarios.}
  \label{fig:deadline_vs_sjf_scatter}
\end{figure}

\begin{figure}[t]
  \centering
  \includegraphics[width=\linewidth]{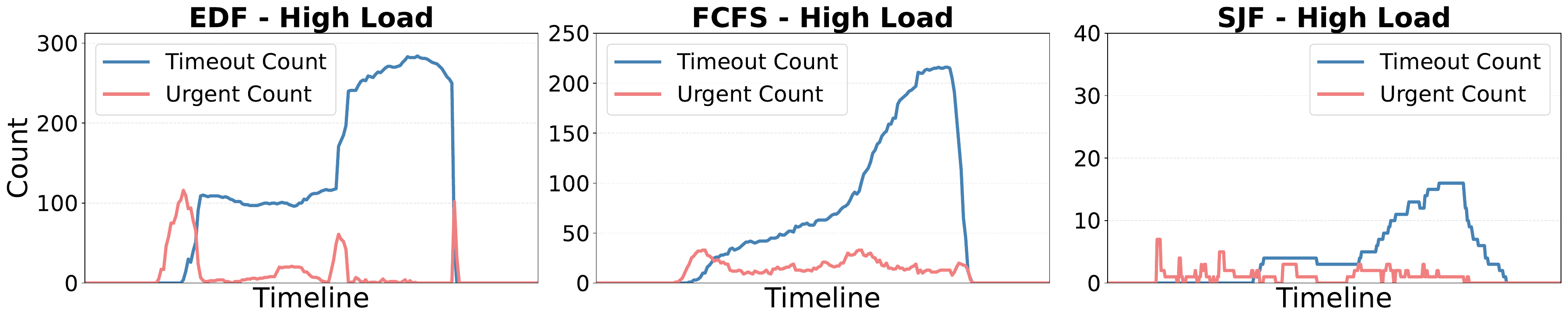}
  \caption{Timeline of urgent and timed-out requests under high load for different scheduling policies.}
  \label{fig:deadline_vs_sjf_timeline}
\end{figure}

\noindent \textbf{Performance of Individual Scheduling Policies.}
Figures~\ref{fig:deadline_vs_sjf_same_token_budget} and~\ref{fig:deadline_vs_sjf_same_batch_size} compare scheduling policies under fixed batch size and token budget on ShareGPT~\cite{sharegpt} and Azure~\cite{azure_trace}. EDF outperforms SJF under low load, but drops sharply beyond a certain load threshold. This occurs because EDF optimistically allocates computational resources to the most urgent requests, assuming all can be completed. Under high load, this assumption fails as many requests become urgent simultaneously, leading to widespread timeouts.

We further provide an in-depth analysis of EDF and SJF under both low and high load conditions. Figure~\ref{fig:deadline_vs_sjf_scatter} shows TTFT/TPOT distributions under low load. While both policies achieve high overall SLO attainment, SJF's bias toward short requests delays longer ones, causing some SLO violations. In contrast, EDF explicitly considers deadlines, preventing starvation and ensuring that all requests meet their TTFT SLOs. Figure~\ref{fig:deadline_vs_sjf_timeline} shows the counts of urgent (approaching deadline) and timed-out (exceeding deadline) requests under high load. EDF causes a sudden
accumulation of urgent requests during peak load, followed by a synchronized surge in timeouts, whereas SJF mitigates long-request blocking and keeps urgent and timeout counts more stable.

\begin{figure}[t]
  \centering
  \includegraphics[width=\linewidth]{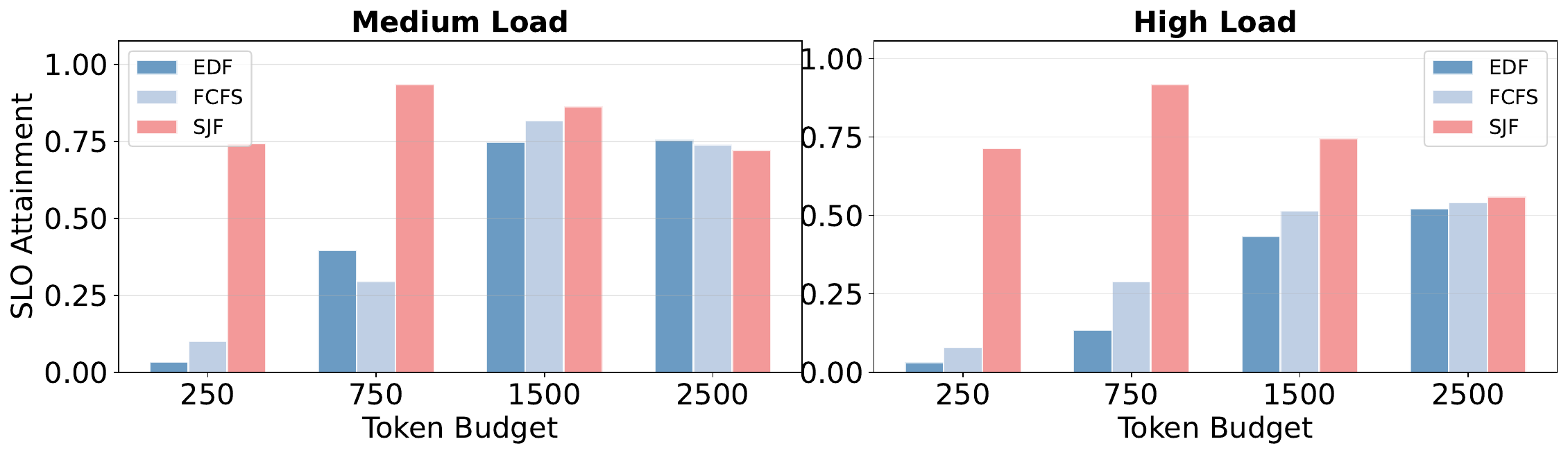}
  \caption{Performance of policies under varying budgets.}
  \label{fig:batch_token_budget_scale}
\end{figure}

\noindent \textbf{Impact of Batch Capacity.}
Figure~\ref{fig:batch_token_budget_scale} shows how different scheduling policies perform across load levels and batch capacity settings. As the token budget increases, EDF steadily improves and eventually stabilizes in SLO attainment. SJF initially improves but then degrades with further budget increases. FCFS behaves differently depending on load: under medium load, its SLO attainment first rises then falls; under high load, it resembles EDF’s trend.
These observations indicate that different scheduling policies have distinct preferences for batch capacity, and this preferred capacity can also fluctuate with changes in the system load. Consequently, a static and fixed batch capacity configuration appears insufficient to accommodate diverse scheduling strategies or adapt to real-world workload dynamics.

\subsection{Hierarchical Memory Management}
\label{sec:memory_manage}

\begin{figure}[t]
  \centering
  \includegraphics[width=\linewidth]{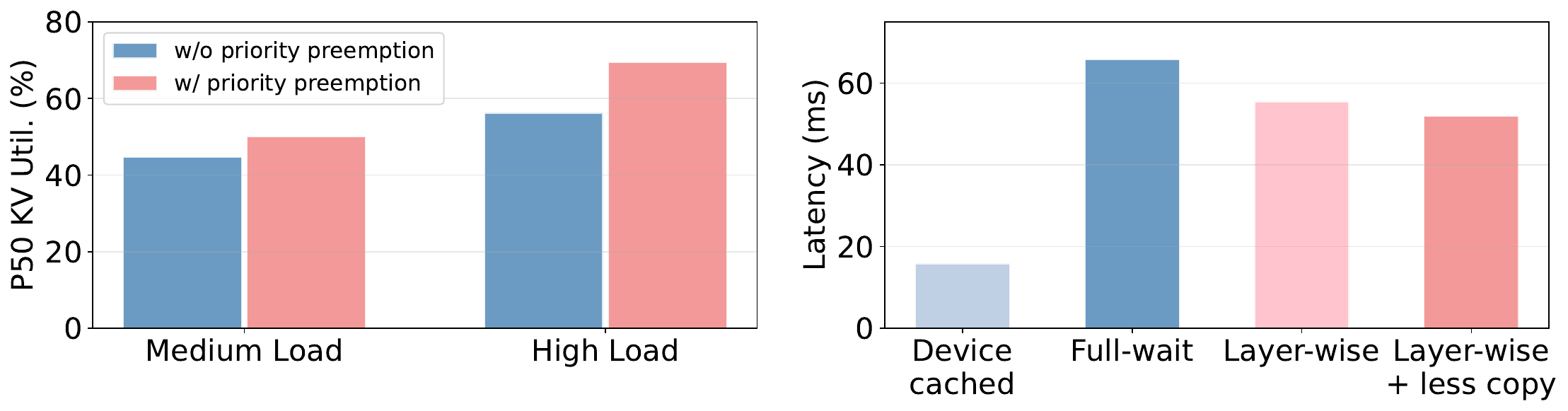}
  \caption{
Left: priority preemption increases p50 KV cache utilization on QwenTrace~\cite{qwen_trace}.
Right: forward latency for a full batch of 4096-token sequences with different modes.
}
  \label{fig:kv_cache_utilization_latency}
\end{figure}

Under traffic bursts and long-output scenarios, KV cache usage grows rapidly. This challenge is further amplified in multi-priority serving, where frequent preemption by high-priority requests can significantly worsen memory inefficiency. 
For instance, low-priority requests may be preempted by later high-priority requests even during the decode phase or after partial chunked prefill, resulting in wasted KV cache.
As shown in the left side of Figure~\ref{fig:kv_cache_utilization_latency}, priority preemption increases KV cache pressure and the effect becomes more pronounced under high load.
The combination of these factors can quickly exhaust device memory, making KV block eviction unavoidable.
Existing approaches~\cite{fastserve, chen2025tokenflow} typically evict blocks and offload them to host memory to avoid extensive recomputation later.
However, several key issues still remain:
(1) host-device block transfers introduce additional latency, requiring careful control of transfer timing and volume to avoid interfering with the critical compute path. As shown in the right panel of Figure~\ref{fig:kv_cache_utilization_latency}, waiting for a full host-to-device reload significantly increases forward latency. Although layer-wise overlap~\cite{xie2025strata} can reduce this delay, further limiting copied blocks can also reduce transfer interference with the main computation path, motivating adaptive block-copy control.
(2) block eviction and reload change request readiness, which couples memory management with scheduling decisions and alters batch formation. This calls for fine-grained coordination between block management and the scheduler.

\begin{figure}[t]
  \centering
  \includegraphics[width=\linewidth]{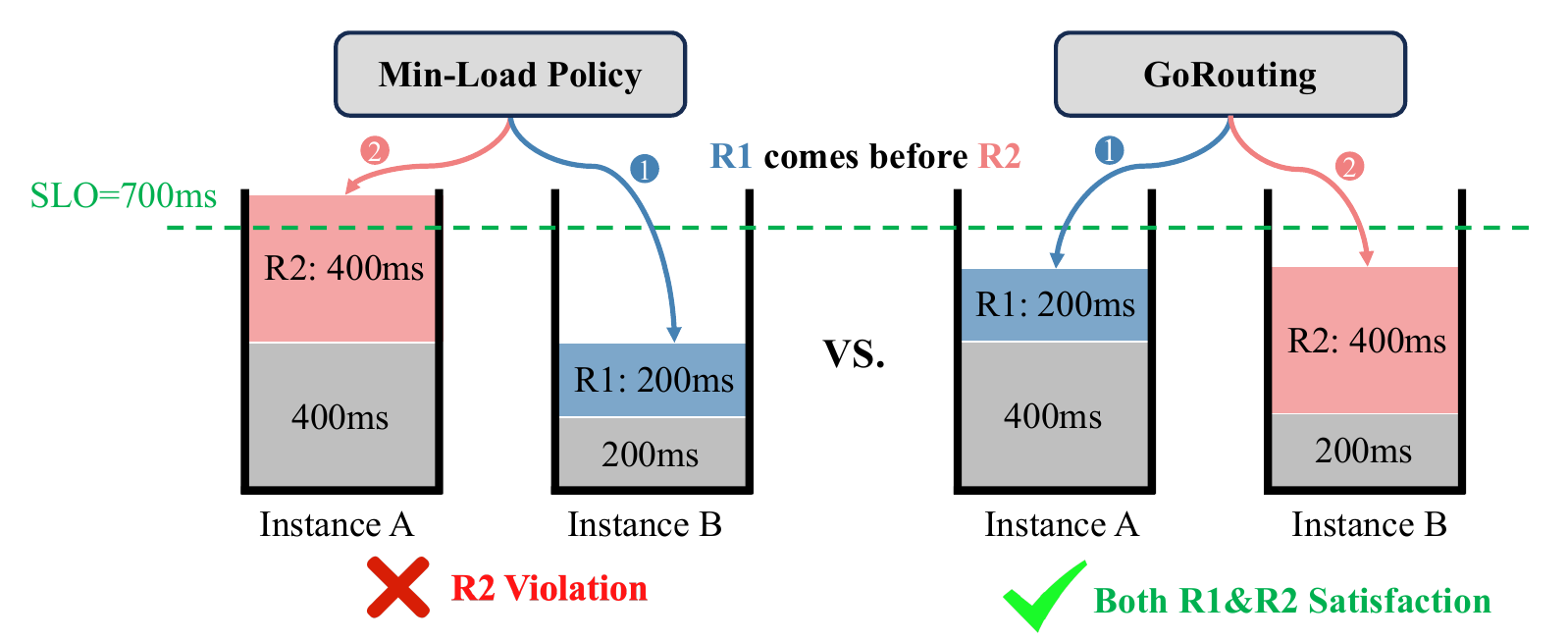}
  \caption{A toy example of the over-balancing issue.}
  \label{fig:overbalance_issue}
  
\end{figure}

\subsection{Limitations of Existing Global Schedulers}
\label{sec:limitation_global}

Existing global schedulers often use least-load dispatching~\cite{vllm, dynamo, xllm} to balance workloads.
However, given fluctuating request arrivals and varying lengths, strict load balancing can be suboptimal: it may fragment service capacity across instances and hinder SLO attainment for future high-priority or long requests.
As illustrated in Figure~\ref{fig:overbalance_issue}, when the longer request R2 arrives shortly after R1, a Min‑Load policy dispatches R1 to the less‑loaded Instance B to balance load instantly, leaving no instance with sufficient slack for R2’s SLO.
Our SLO‑aware strategy instead dispatches R1 to the moderately loaded Instance A, which still meets R1’s SLO while preserving capacity on Instance B for R2. Although R1’s TTFT increases slightly, both requests meet their deadlines.
This demonstrates the need to move from pure load‑balancing to a capacity‑aware, SLO‑driven policy that accounts for request length, deadline, and local scheduler behavior.

\section{Design}
\label{sec:design}
\begin{figure}[t]
  \centering
  \includegraphics[width=\linewidth]{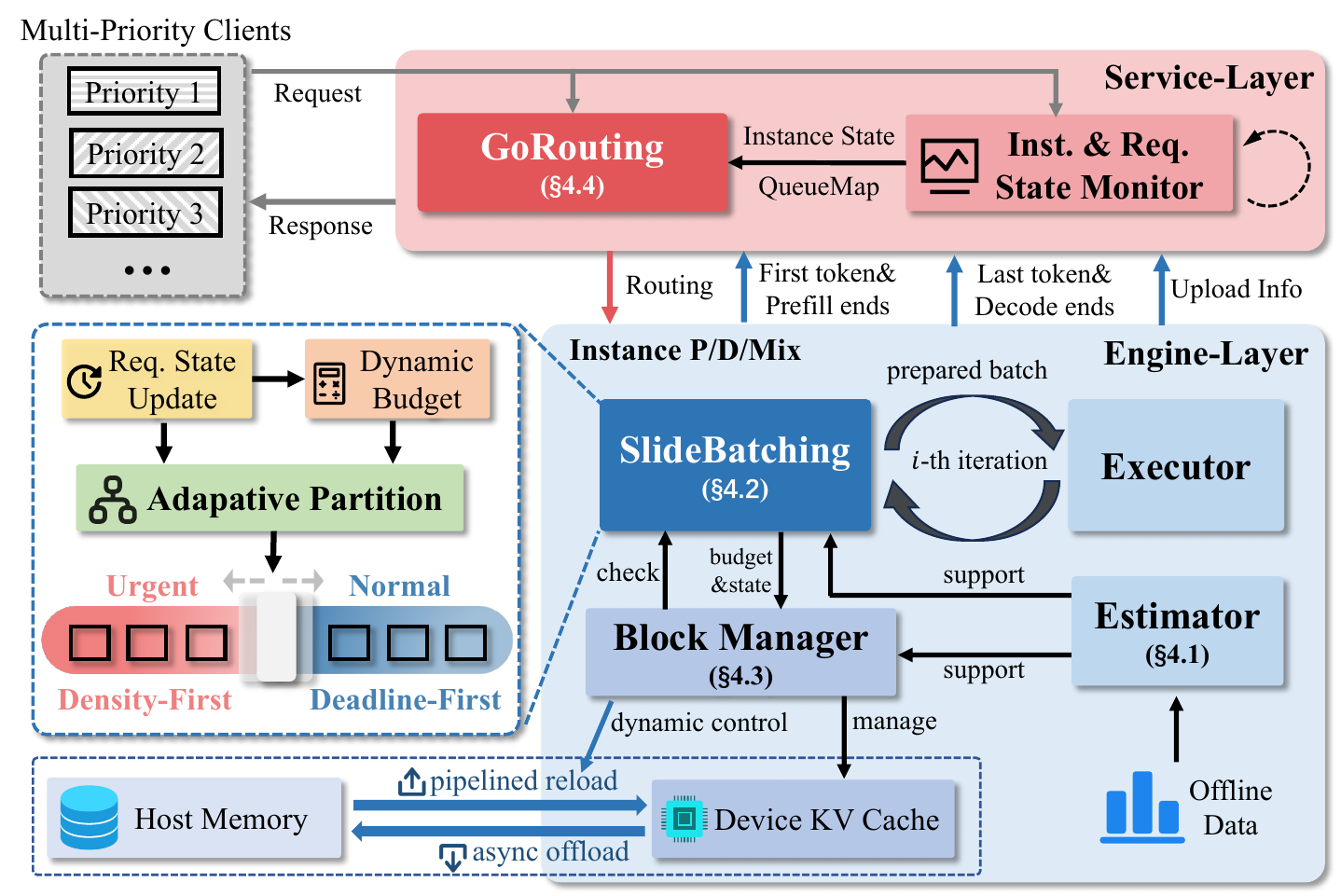}
  \caption{The overall framework of \systemname{}.}
  \label{fig:framework}
\end{figure}

The overall architecture of \systemname{} is shown in Figure~\ref{fig:framework}.
\systemname{} comprises several primary components: \localsched{} (\textsection\ref{sec:local_sched}) and efficient memory management (\textsection\ref{sec:block_manage}) at the engine layer and \globalsched{} (\textsection\ref{sec:global_sched}) at the service layer.
\systemname{} targets both PD co-located and PD disaggregated deployments since both modes are widely adopted by modern inference engines~\cite{vllm,liu2025xllm}.
In PD co-location, the prefill and decode phases run on the same instance, whereas in PD disaggregation, prefill and decode run on different instances.


\subsection{Batch Latency Estimator}
\label{sec:batch_estimator}
The estimation of batch execution time is crucial for batch scheduling. The batch execution time can be decomposed into constant overhead (e.g., kernel launch and input/output processing), computation time, and memory access time. In general scenarios, a batch may contain both prefill and decode requests, where prefill requests are typically compute-intensive~\cite{sarathi-serve}, dominated by linear and attention operations, whereas decode requests are typically memory-bound~\cite{distserve}. Hence, we formulate separate linear regression models for prefill and decode requests as follows:
\begin{align}
T_{pd}(r) &= \tilde{T}_{pd}(r) + t_c,\ \tilde{T}_{pd}(r) = \begin{cases}
\tilde{T}_p(r) & ,r\ \text{is prefill}\\
\tilde{T}_d(r) & ,r\ \text{is decode}\\
 \end{cases}, \\
\tilde{T}_p(r) &= a_p \cdot l_q(r)^2 + b_p \cdot l_q(r) \cdot l_{kv}(r) + c_p \cdot l_q(r), \\
\tilde{T}_d(r) &= a_d \cdot l_{kv}(r) + b_d \cdot 1, 
\end{align}
where $T_{pd}(r)$ denotes the estimated latency for request $r$ including a constant overhead $t_c$, while $\tilde{T}_{pd}(r)$ represents the core computational latency estimate excluding this fixed overhead. $\{a_p, b_p, c_p, a_d, b_d\}$ are trainable parameters. $l_{kv}(r)$ denotes the KV cache length, and $l_{q}(r)$ is the number of tokens processed in the current forward pass. This design is directly compatible with chunked prefill~\cite{patel2024splitwise} and prefix caching~\cite{zheng2024sglang} features.
Then the execution time for a batch $B$ can be estimated as:
\begin{equation}
T_{pd}(B) = T_{pd}(B_p \cup B_d) = \sum_{r \in B_p} \tilde{T}_p(r) + \sum_{r \in B_d} \tilde{T}_d(r) + t_c.
\end{equation}

We leverage offline-generated profiling batch data to train the models and similarly construct an evaluation dataset. The evaluation results show that the Mean Absolute Percentage Error (MAPE) remains stable at approximately 4.5\%.



\subsection{Local Scheduler: SlideBatching}
\label{sec:local_sched}

The key to batch scheduling lies in determining the batch capacity and the order of requests to admit. Our core design principle is: \textit{when possible, satisfy all request deadlines to capture full system gain; when current load cannot meet the deadlines of all requests, prioritize high-priority requests to maximize overall gain}. To this end, we propose the \localsched{} algorithm that effectively balances request latency and priority.

\noindent \textbf{Algorithm.} 
While batch capacity is typically defined by token budget or batch size, thanks to the latency estimator introduced in \textsection\ref{sec:batch_estimator}, we can directly transform token- or sequence-level budgets into a time-level \textit{latency budget}, comparable to request deadlines. Alg.~\ref{alg:batch_policy} first sets this budget to the smallest remaining deadline among queued requests (line~6), ensuring no request misses its deadline in the current batch. To avoid overly small batches under high load, a lower bound $\eta$ is enforced (line~7). The core of the algorithm is the request ordering strategy (line~13): we prioritize \textsc{Urgent} requests, and within the \textsc{Urgent} group, we schedule requests in descending order of their \textit{density} (line~5). The remaining requests are scheduled in ascending order of their remaining time to deadline ($r.\text{remain}$). The numbers of \textsc{Urgent} and \textsc{Normal} requests vary dynamically with the queue load, and the boundary between them \textquotedblleft slides\textquotedblright~accordingly.

\begin{algorithm}[t]
    \caption{SlideBatching Algorithm}
    \label{alg:batch_policy}
    \renewcommand{\algorithmicrequire}{\textbf{Input:}}
    \renewcommand{\algorithmicensure}{\textbf{Output:}}
    \begin{algorithmic}[1]
        \Require the request queue $Q$
        \Ensure prepared batch of requests $B$
        \State $t_{min} \gets \infty$
        \For{$r$ in $Q$} \Comment{\texttt{Update request metric}}
            \State $r.exec \gets \tilde{T}_{pd}(r)$
            \State $r.remain \gets deadline_{r,\ r.len} - r.elapsed\_time$
            \State $r.density \gets \frac{w_r(r.len)}{r.exec}$
            \State $t_{min} \gets \min(t_{min}, r.remain)$
        \EndFor
        \State $t_{budget}\gets \max(t_{min} ,\eta)$
        \For{$r$ in $Q$} \Comment{\texttt{Determine requests' urgency}}
            \If{$r.remain < \gamma\cdot \phi(r,Q)$}
                \State $r.state \gets \textsc{Urgent}$
            \Else
                \State $r.state \gets \textsc{Normal}$
            \EndIf
        \EndFor
        \State Sort $Q$ first by $r.state$, then sort \textsc{Urgent} requests in descending order of $r.density$ and sort \textsc{Normal} requests in ascending order of $r.remain$.
        \State $B_{\text{copy}} \gets \textsc{GetCopyBudget}(Q, t_{budget})$
        \State $t_{batch} \gets t_c$
        \While{$Q$ \textbf{and} $t_{batch} < t_{budget}$ \textbf{and} $memory\ is\ enough$}
            \State $r \gets Q.\text{pop}()$
            \State $t,c \gets \textsc{GetMaxChunk}(r, t_{budget}-t_{batch})$
            \If{$\neg\textsc{SatisfyCopyCondition}(c, r, B_{\text{copy}})$}
                \State \textbf{continue}
            \EndIf
            \State $B_{\text{copy}} \gets \textsc{ConsumeCopyBudget}(r, B_{\text{copy}})$
            \State $t_{batch} \gets t_{batch} + t$
            \State $B \gets B \cup \{r\}$
        \EndWhile
        
        \State \Return $B$
    \end{algorithmic}
\end{algorithm}

\noindent \textbf{Adaptive Urgency Partition.}
The algorithm dynamically diagnoses request urgency (lines~9--13) to prioritize \textsc{Urgent} requests that are at risk of missing deadlines.
A request is considered \textsc{Urgent} if it is likely to miss its deadline under the current load.
This assessment requires a load‑judgment function $\phi(r,Q)$, which can be compared with the request's remaining time $r.\text{remain}$ to determine its urgency. Since a request excluded from the current batch may still be served in a subsequent batch, $\phi$ should estimate whether the request can be completed in any future batch.

In \textit{PD co‑location}, we approximate the latency budget of future batches using the current $t_{budget}$. This approximation is reasonable because, when the queue cannot be fully served, the algorithm tends to saturate $t_{\text{budget}}$ via chunked prefill (line~21). Consequently, each batch contains requests that either just meet or miss deadlines. If such a request is still running (implying it has entered the decoding phase), the minimum remaining time to deadline in the next batch will be shorter than its $\text{TPOT}_{\text{SLO}}^r$. Given the lower bound $\eta$, $t_{budget}$ fluctuates within $[\eta, \max(\text{TPOT}_{\text{SLO}}^r)]$. Since requests' TPOT SLOs are relatively small, successive budgets remain relatively stable. We then adopt a worst‑case implementation for $\phi(r,Q)$ that statically places every request $r$ at the end of the queue, resulting in a request‑agnostic formula:
\begin{equation}
    \phi(Q) = \frac{t_{budget}}{t_{budget} - t_c} \sum_{r \in Q} r.\text{exec},
    \label{eq:phi_a}
\end{equation}
where $t_c$ denotes the constant per‑batch overhead introduced in \textsection\ref{sec:batch_estimator} and the term $\frac{\sum_{r \in Q} r.\text{exec}}{t_{\text{budget}} - t_c}$ represents the number of batch steps required to process the entire request queue $Q$.




Additionally, we also introduce an \textit{aggressiveness coefficient} $\gamma$ (line~10), which can be adjusted manually. A larger $\gamma$ shifts more requests towards the \textsc{Urgent} side, favoring density‑first scheduling pessimistically to capture short‑term gain, whereas a smaller $\gamma$ shifts more requests toward the \textsc{Normal} side, favoring deadline‑first scheduling optimistically in pursuit of long‑term gain.

In \textit{PD disaggregation}, we only schedule the prefill‑only instance, as decode requests are interference‑free and typically batched together~\cite{pdagg_ornot}. Given that prefill execution times are relatively long, we also adopt a worst‑case estimation which leads to a simplified load‑judgment function: $\phi_p(Q) = \sum_{r \in Q} r.\text{exec} + |Q| \cdot t_c,$ where each batch is assumed to contain only one request.

\noindent \textbf{Analysis.}
The algorithm's behavior adapts to the system's actual load through its sliding boundary mechanism.
(1) \textbf{Low Load:} The latency budget accommodates all queued requests, rendering the specific ordering strategy negligible in its impact.
(2) \textbf{Medium load:} A subset of requests is classified as \textsc{Urgent}, but the batch capacity is still sufficient to accommodate all urgent requests along with some normal ones. The scheduling policy remains predominantly deadline-first, prioritizing requests with the earliest deadlines.
(3) \textbf{High load:} The number of urgent requests increases sharply, exceeding the batch capacity, which means that only a subset of urgent requests can be admitted into the current batch. As long as $t_{min}$ is greater than threshold $\eta$, each admitted request is expected to meet its deadline and yield its gain. Moreover, urgent requests not admitted to the current batch are likely to miss their deadlines under high load or due to preemption. 
In this context, local batch formation can be heuristically viewed as a \textit{fractional-knapsack} problem~\cite{salkin1975knapsack}: the latency budget $t_{\text{budget}}$ acts as capacity, $r.\text{exec}$ as item size, and token-level gain $w_r(i)$ as item value. 
With chunked prefill, requests can be approximated as divisible items, which motivates the optimal density-first greedy policy for constructing the current iteration's batch. 

\noindent \textbf{Starvation Prevention.}
While our scheduler accounts for SLO deadlines, sustained overload may still allow high-priority traffic to monopolize compute resources, resulting in excessive delays for low-priority requests.
We therefore support a configurable anti-starvation rule: once a request's waiting time exceeds an adjustable starvation threshold $\tau$, it is marked as starving and temporarily promoted to the head of $Q$ for next-batch admission.

\subsection{Efficient Block Management}
\label{sec:block_manage}
Under device memory pressure, especially with frequent preemption from high-priority traffic, KV cache eviction becomes unavoidable. The following design addresses two concerns: (i) which requests' blocks to evict under memory pressure, and (ii) how to minimize the performance impact of eviction and resume.

\noindent \textbf{Eviction Policy.} When device memory is saturated, we preferentially evict KV cache blocks of requests near the tail of sorted $Q$, except those whose waiting time is close to the starvation threshold. These requests are unlikely to be scheduled in the near term, so their blocks offer limited short-term value. Instead of dropping evicted blocks, we spill them to host memory and restore them upon rescheduling, preserving reuse and avoiding redundant recomputation.

\noindent \textbf{Asynchronous Offloading.} 
Offloading KV cache from device to host memory introduces latency, and critically, the device memory occupied by the blocks cannot be reclaimed until the transfer completes.
To mitigate this overhead, we spawn a separate thread to asynchronously copy accumulated blocks to host memory, triggering a copy operation for every $n_{\text{off}}^r$ newly generated blocks of request $r$.
We make the offloading threshold $n_{\text{off}}^r$ priority-aware by assigning lower-priority requests smaller thresholds to increase copy frequency, as they are more likely to be preempted by later high-priority requests.
This proactive approach overlaps transfer with computation and effectively avoids on-demand synchronization: when a request's eviction eventually occurs, we can directly evict all its device blocks and discard the pending transfer.

\noindent \textbf{Pipelined Reloading.}
When an evicted request is readmitted to a batch, its KV cache must be restored to device memory before computation can proceed.
Instead of synchronously reloading all blocks upfront, we exploit the inherent layer-by-layer execution characteristic of current LLMs and implement \textit{pipelined asynchronous uploading}~\cite{xie2025strata}: while the model computes layer \(i\), the system asynchronously loads the KV cache blocks for layer \(i+1\) from host to device memory. 
Nevertheless, a critical question remains: \textit{how many blocks should be uploaded per batch?} Uploading too few blocks underutilizes device compute capacity, whereas uploading too many may prolong the critical path and risk violating SLOs. This trade-off motivates our following adaptive copy-budget control mechanism.

\noindent \textbf{Adaptive Copy-Budget Control.}
To effectively overlap computation and data transfer across layers, we dynamically determine the optimal number of blocks to upload. Given that current LLMs typically exhibit consistent layer structures, the computation and transfer times across layers are approximately uniform. We maintain the number of blocks resident in host memory for each request $r$, denoted as \(b_{\text{host}}^r\). Based on this information, we estimate two key metrics: (1) \textit{Minimum forward latency} \(T_{fwd}^{min}\): the estimated forward time assuming host-resident blocks for all requests have been copied to the device; (2) \textit{Maximum transfer time} \(T_{trans}^{max}\): the time needed to copy all missing blocks, calculated as \( T_{trans}^{max} = \sum_{r} b_{\text{miss}}^r \times t_{\text{block}}\), where \( b_{\text{miss}}^r = \max(0,b_{\text{host}}^r-b_{\text{device}}^r) \) and \(t_{\text{block}}\) denotes the profiled copy time per block.

We then determine the maximum number of blocks to copy \(B_{\text{copy}}\) using the following decision procedure:
(1) \(T_{fwd}^{min} > t_{budget}\): As the batch inference time is dominated by the latency budget, we have \(B_{\text{copy}} = \left\lfloor \frac{t_{budget}}{t_{\text{block}}} \right\rfloor\).
(2) \(T_{fwd}^{min} < t_{budget}\): The batch inference time is primarily determined by the execution time of all requests in the queue. We further analyze the following two subcases:
(i) \(T_{fwd}^{min} > T_{trans}^{max}\): Computation dominates the total time. We can safely copy all missing blocks (\(B_{\text{copy}} = B_{\text{missing}}\)).
(ii) \(T_{fwd}^{min} < T_{trans}^{max}\): Transfer time risks becoming the bottleneck. As \(B_{\text{copy}}\) increases, the actual batch latency decreases monotonically (approaching \(T_{fwd}^{min}\)), while the transfer time increases monotonically (approaching \(T_{trans}^{max}\)). By monotonicity, we apply binary search to find the largest feasible \(B_{\text{copy}}\) such that the estimated transfer time does not exceed the estimated batch latency.


\noindent \textbf{Put it Together with \localsched{}.} 
\localsched{} and Block Management operate in a closed loop, as reflected in Alg.~\ref{alg:batch_policy}. In each scheduling round, \localsched{} first exports the ordered request queue \(Q\) and the current latency budget \(t_{budget}\). Block Management then uses this information to compute the host-to-device copy budget \(B_{\text{copy}}\) following the process above.
Given \(B_{\text{copy}}\), \localsched{} consumes this budget in queue order at request granularity. For request \(r\), if the remaining budget can cover all missing blocks \(b_{\text{miss}}^r\), we copy all of them; otherwise, we consider partial copying with residual budget \(B_{\text{rem}}\). 
Let \(l_{\text{comp}}^r\) denote the number of tokens that can still be computed for request \(r\) after partial copying in the current round (it may be capped by \(t_{budget}\) ). Let \(s_{\text{blk}}\) denote the token capacity per block. We enable partial copying only when it yields sufficient effective progress: either \(l_{\text{comp}}^r\) reaches the maximum computable-token limit for request \(r\) in this round, or the ratio \(\frac{l_{\text{comp}}^r}{(b_{\text{miss}}^r - B_{\text{rem}})\cdot s_{\text{blk}}}\) exceeds a threshold \(\beta\), where \(\beta > 1\); otherwise, \(r\) is skipped in this round and the scheduler proceeds to the next request. This policy helps keep transfer from becoming the critical path while respecting \localsched{}'s admission order.

\subsection{Global Scheduler: GoRouting}
\label{sec:global_sched}

\begin{algorithm}[t]

    \caption{GoRouting Algorithm in PD Disagg.}
    \label{alg:request_dispatch_policy}
    \renewcommand{\algorithmicrequire}{\textbf{Input:}}
    \renewcommand{\algorithmicensure}{\textbf{Output:}}
    \begin{algorithmic}[1]
\Require prefill instance pool $P$, decode instance pool $D$, request $req$, per-instance prefill queues $\{Q^{\text{pre}}_p\}_{p\in P}$ maintained for instances
\Ensure Selected instance pair $(p\_inst, d\_inst)$

\State $\Delta_{\max} \gets -\infty$

\For{$p$ in $P$} \Comment{\texttt{Estimate each instance's gain}}

    \State $pre\_gain \gets \textsc{EstimateGain}(Q^{\text{pre}}_p)$
    \State $post\_gain \gets \textsc{EstimateGain}(Q^{\text{pre}}_p\cup \{req\})$
    \Statex \Comment{\texttt{Based on specific local scheduler}}
    
    \State $\Delta_p = \mathit{post\_gain} - \mathit{pre\_gain}$
    \State $\Delta_{\max} \gets \max(\Delta_{\max}, \Delta_p)$
\EndFor

\State $\mathit{C} \gets \{ p \in P \mid \Delta_p \ge \alpha \cdot \Delta_{\max} \}$ 
\Comment{\texttt{Candidates}}

\If{$\Delta_{\max} > 0$}
    \State $L \gets \{ p \in \mathit{C} \mid \textsc{EstimateExec}(p) < \mu \cdot TTFT_{SLO} \}$
    \State $H \gets  \{ p \in \mathit{C} \mid \textsc{EstimateExec}(p,req) > \lambda \cdot TTFT_{SLO} \}$
    \If{$L \neq \emptyset$}
        \State $p\_inst \gets \arg\min_{p \in L} \textsc{EstimateExec}(p)$
    \ElsIf{$C-H \neq \emptyset$}
        \State $p\_inst \gets \arg\max_{p \in \mathit{C-H}} \textsc{EstimateExec}(p)$
    \Else
        \State $p\_inst \gets \arg\min_{p \in \mathit{C}} \textsc{EstimateExec}(p)$
    \EndIf

\Else
    \State $p\_inst \gets \textsc{SelectMinLoad}(P)$ \Comment{\texttt{Fallback}}
\EndIf
\State $d\_inst \gets \textsc{SelectMaxFree}(D)$
\State \Return $(p\_inst, d\_inst)$

    \end{algorithmic}
\end{algorithm}

We present this part in three steps: unified instance-state monitoring, instance selection for PD disaggregation, and extension of the same policy to PD co-location.

\noindent \textbf{Instance State Monitoring and Update.}
\label{sec:inst_monitor}
To accurately track local instance state for global scheduling while minimizing communication and update overhead, the global scheduler uses a unified monitoring framework across both deployment modes and maintains several lightweight states: the per-instance prefill queue \( Q^{\text{pre}}_p \), the decode counter \( n_d \), and the number of free blocks \( b_{f} \).
\( b_f \) is collected periodically from per-instance reports of available blocks.
Other states are updated in an event-driven manner: when request \( r \) is dispatched to instance \( p \), \( Q^{\text{pre}}_p \leftarrow Q^{\text{pre}}_p \cup \{r\} \); when prefill finishes, \( Q^{\text{pre}}_p \leftarrow Q^{\text{pre}}_p \setminus \{r\} \) and it increments \( n_d \); when the request completes, it decrements \( n_d \).
In \textbf{PD disaggregation}, \textit{prefill} instances mainly track \( Q^{\text{pre}}_p \), whereas \textit{decode} instances mainly track \( n_d \). In \textbf{PD co-location}, each \textit{mix} instance tracks both \( Q^{\text{pre}}_p \) and \( n_d \).

However, event-driven updates can still introduce state staleness because \( Q^{\text{pre}}_p \) is updated only at dispatch and prefill-completion events, while prefill execution is relatively long. As a result, the scheduler's view may lag behind actual in-flight prefill progress, leading to biased remaining-time estimation and suboptimal dispatch decisions. To compensate, the global scheduler records a timestamp \( ts_p \) at first insertion and each removal from \( Q^{\text{pre}}_p \). When estimating execution time on instance \( p \), it first subtracts the elapsed interval \( ts_{\text{curr}} - ts_p \) when \( Q^{\text{pre}}_p \neq \emptyset \), and then applies a formula similar to the load-judgment function \(\phi\) (Eq.~\ref{eq:phi_a}).

\begin{figure*}[t]
  \centering
  \includegraphics[width=\linewidth]{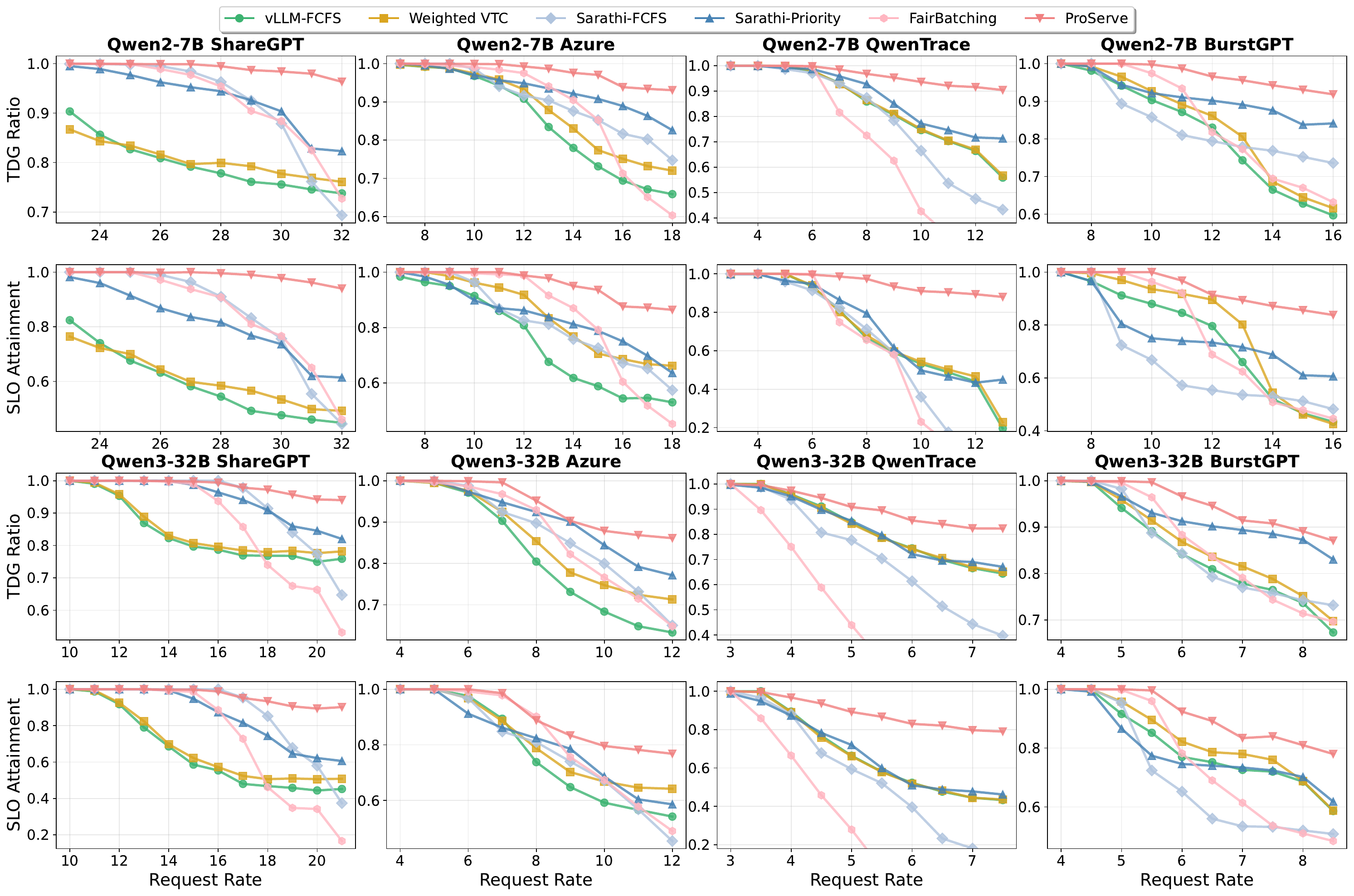}
  \caption{Single-node performance of different methods across different datasets and models.}
  \label{fig:main_results}
\end{figure*}

\noindent \textbf{Instance Selection in PD Disaggregation.}
\label{sec:pd_disagg_dispatch}
The request-dispatching algorithm is detailed in Alg.~\ref{alg:request_dispatch_policy}.
For each prefill instance $p$, we invoke \textsc{EstimateGain} to evaluate the gain before and after admitting the incoming request (lines~2--5). \textsc{EstimateGain} is instantiated by the local scheduler: it estimates request execution times under the scheduler's queueing policy, compares them against the corresponding remaining TTFT budgets, and aggregates the resulting per-request gains.
To avoid over-committing to a single instance when several are similarly good, we introduce a candidate set $C = \{p \in P \mid \Delta_p \ge \alpha \cdot \Delta_{\max}\}$ (line~7), where $\Delta_{\max} = \max_{p \in P} \Delta_p$ is the maximum incremental gain across instances. Thus, when $\Delta_{\max} > 0$, every instance in $C$ can satisfy the current request's SLO.

Furthermore, as discussed in \textsection\ref{sec:limitation_global}, purely balanced dispatching is less robust to request-length fluctuations and can cause later long requests to miss SLOs. We therefore use a dual-threshold, capability-aware policy (lines~8--16). Let $L$ and $H$ denote the light-load and heavy-load subsets of $C$, defined by thresholds $\mu$ and $\lambda$, respectively. The final prefill-instance selection is: \textbf{(1) If $L \neq \emptyset$}: select the most idle instance in $L$ to avoid under-utilization. \textbf{(2) If $C \setminus H = \emptyset$} (all candidates are heavily loaded), fall back to load balancing and select the least-loaded instance in $C$ to avoid overload.
\textbf{(3) Otherwise}, select the relatively heaviest instance in $C \setminus H$. Although this may increase the TTFT of the dispatched request, the properties of $C$ guarantee that the request still meets its TTFT SLO and achieves its gain, while reserving capacity on lighter-loaded instances for future potentially long or high-priority requests.

For decode instances, execution is decoupled from prefill~\cite{distserve}. Since decode is typically memory-bound, we select the instance with the largest number of free blocks \( b_f \).

\noindent \textbf{Extension to Instance Selection in PD Co‑location.}
Under PD co-location, each request stays on the same instance across both prefill and decode, coupling the two phases and making gain estimation more involved. For tractability, we assume decode requests are always included in the batch. This assumption is reasonable for our SlideBatching scheduler, since decode execution is short and TPOT SLOs are typically tight, which naturally prioritizes decode latency. It is also consistent with Sarathi-Serve, which strictly prioritizes decode scheduling. Accordingly, we evaluate with both local schedulers in \textsection\ref{sec:main_result}. We therefore compute gain from the prefill side via \textsc{EstimateGain}, allowing Alg.~\ref{alg:request_dispatch_policy} to be reused with minimal changes.
The adjustment is that, when estimating TTFT time, we add an extra decode overhead \(\hat{t}_d(n_d)\), as detailed in Appendix~\ref{gorouting_details}. In addition, to further guarantee decode latency, if \(\hat{t}_d(n_d + |Q^{\text{pre}}|)\) approaches the TPOT SLO, we exclude that instance from the candidate set.





\section{Evaluation}

\subsection{Experimental Setup}
\noindent \textbf{Datasets and Workloads.}
We evaluate our method on four open‑source datasets: ShareGPT~\cite{sharegpt}, Azure~\cite{azure_trace}, BurstGPT~\cite{wang2025burstgpt}, and QwenTrace~\cite{qwen_trace}.
For datasets with real timestamps (Azure, QwenTrace, and BurstGPT), we adopt a commonly used scaling method~\cite{arrow, mooncake, wang2025burstgpt}. This approach expands the timestamps according to a pre‑defined overall request rate, and then replays the requests following the scaled intervals.
For datasets lacking real timestamps (ShareGPT), we employ a Poisson distribution to simulate the request arrival pattern.
Additionally, we also include our proprietary industrial dataset, which will be described in \textsection\ref{sec:cluster}.

\noindent \textbf{Testbed and Models.}
We deploy \systemname{} on the recently open-sourced and high-performance xLLM~\cite{liu2025xllm} inference framework, with each server equipped with 16 Ascend 910B NPUs, 96 physical CPU cores, and 2~TB of RAM. We select Qwen2-7B~\cite{qwen2} and Qwen3-32B~\cite{qwen3} for evaluation.

\noindent \textbf{Metrics.}
We select the TDG introduced in \textsection\ref{sec:priority_objective} as the gain function to quantify unified service gain for multi-priority requests. We further define the system-level service gain metric as $\text{TDG\_Ratio} = \frac{\sum_r f_{TDG}(r)}{Ideal\_Gain}$, which represents the proportion of captured gain to the total achievable gain. In addition, we report the SLO attainment ratio as an overall latency-performance metric, which is widely adopted in recent work~\cite{arrow, dong2025hydrainfer, huang2025slo, hyperflexis, tang2025scorpio}. A request is considered to have met its SLO only when both its observed TTFT and TPOT are strictly less than the preset SLO thresholds.

\noindent \textbf{Baselines.}
We compare \systemname{} against the following batch scheduling algorithms:
\begin{itemize}
    \item \textbf{vLLM-FCFS}~\cite{vllm}: The default scheduling algorithm in vLLM. It prioritizes prefill requests and employs FCFS.

    \item \textbf{Weighted VTC}~\cite{vtc}: A variant of the VTC algorithm. 
    Weighted VTC assigns different priority weights (analogous to the \textit{nice} values in Linux) to clients, ensuring the ratio of tokens processed approximates the ratio of assigned priority weights.

    \item \textbf{Sarathi-FCFS}~\cite{sarathi-serve}: The scheduler in Sarathi-Serve, which employs chunked prefill. It prioritizes decode requests and uses FCFS within each request type. It uses profiled token budget based on TBT.

    \item \textbf{Sarathi-Priority}: A priority-based extension of Sarathi. It prioritizes decode requests first, followed by those with higher priority, and finally, those that arrived earlier.

    \item \textbf{FairBatching}~\cite{fairbatching}: A recently enhanced EDF scheduling policy. It schedules requests by prioritizing decode sequences nearing their deadlines, followed by prefill sequences, and finally the remaining decode requests.
\end{itemize}
In multi‑node experiments, we use the widely adopted \textbf{MinLoad} strategy as the global scheduler baseline, which dispatches each request to the least‑loaded instance. For a fair comparison, all schedulers are uniformly implemented within the xLLM~\cite{liu2025xllm} framework.

\begin{figure*}[t]
  \centering
  \includegraphics[width=\linewidth]{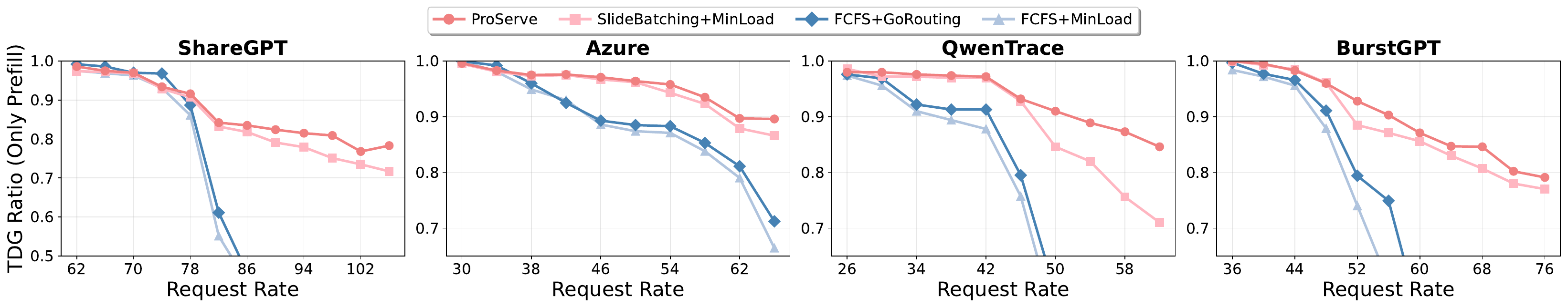}
  \caption{Multi-node performance of different methods across datasets under PD disaggregation.}
  \label{fig:pd_disagg_results}
\end{figure*}

\begin{figure}[t]
  \centering
  \includegraphics[width=\linewidth]{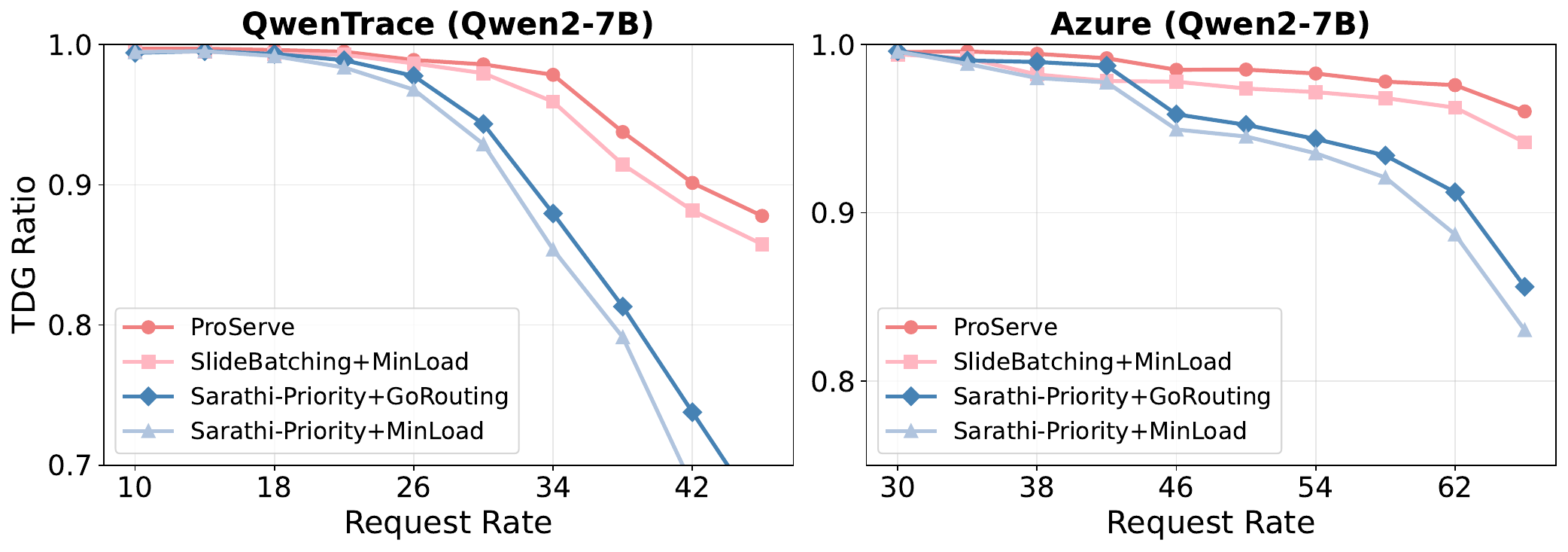}
  \caption{Multi-node performance of different methods across datasets under PD co-location.}
  \label{fig:multi_ins_results}
\end{figure}

\noindent \textbf{Details.}
To simulate a multi‑priority scenario, each request in the dataset is randomly designated as high or low priority with a 50\% probability. In our experiments, the priority weights are fixed at 2 and 1 for high‑ and low‑priority requests, respectively. An analysis of how different priority weight configurations affect our method is provided in Appendix~\ref{sec:priority_scale}. The ratio between the first‑token weight and the decode‑token weight in TDG is configured based on the average ratio of input to output length from the dataset.

\subsection{Main Results} \label{sec:main_result}
\noindent \textbf{Single-Node Performance.}
Figure~\ref{fig:main_results} reports batch scheduling performance under single-node PD co-location and shows that \systemname{} consistently achieves the best TDG and SLO attainment across all tested datasets and models.
Deadline-first strategies (e.g., FairBatching and Sarathi-FCFS) perform well under low load, matching \systemname{} in system gain, but degrade sharply at higher request rates and eventually fall below vLLM-FCFS and Weighted VTC. This aligns with our analysis in \textsection\ref{sec:static_scheduler}.
Although Sarathi-Priority and Weighted VTC are priority-aware, each has a critical limitation: Sarathi-Priority's strict prioritization starves low-priority requests and hurts overall gain, while Weighted VTC focuses on weighted token fairness but ignores SLO constraints, resulting in lower TDG and SLO attainment.

\noindent \textbf{Multi-Node Performance.}
Since the PD-disaggregated setting inherently favors decode requests, the TDG for decode tokens is almost always satisfied in our experiments. We report only the first-token TDG. 
First, as shown in Figure~\ref{fig:pd_disagg_results}, \globalsched{} enhances various local schedulers. While not always selecting the least-loaded node, its SLO-aware dispatch performs comparably under light load. At higher loads, it reserves capacity for future long requests, thereby improving overall TDG.
Second, the improvement with SlideBatching is more pronounced than with GoRouting. This is because GoRouting's effectiveness depends on specific traffic patterns (e.g., the QwenTrace dataset exhibits higher variance in request lengths, leading to more significant gains), whereas SlideBatching adapts better to diverse request arrivals.
Moreover, results are consistent under PD co‑location (Figure~\ref{fig:multi_ins_results}).

\begin{figure}[t]
  \centering
  \includegraphics[width=\linewidth]{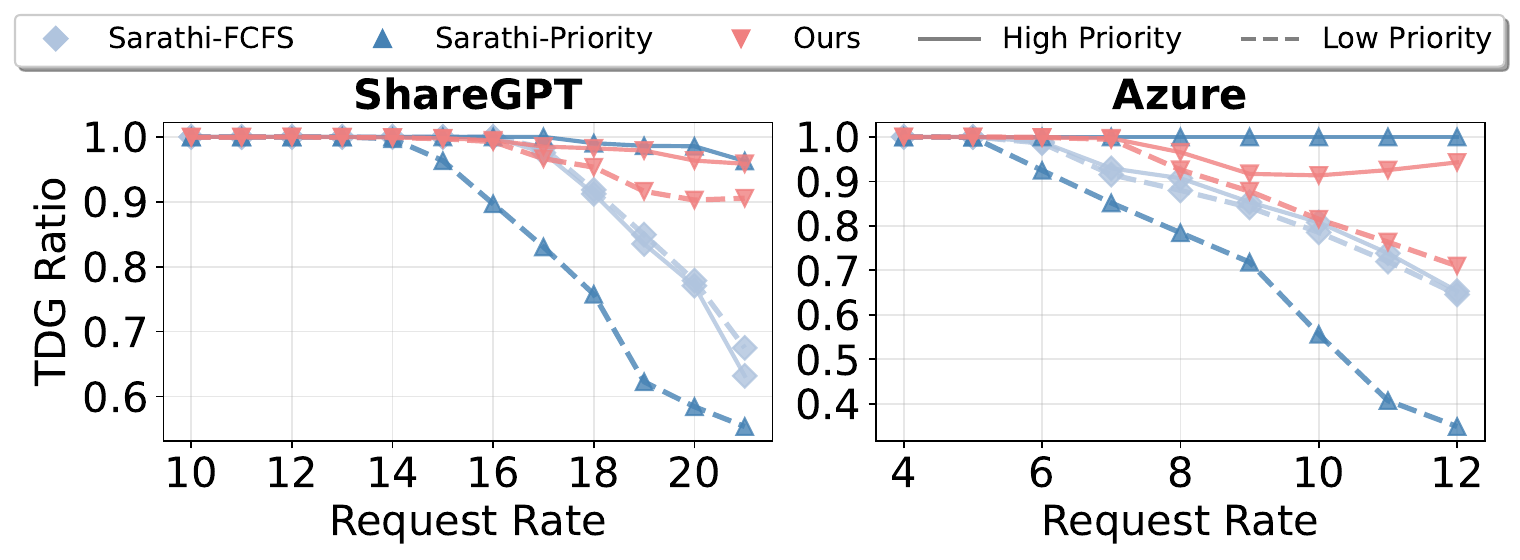}
  \caption{Performance for different priority requests under various scheduling strategies using Qwen3-32B.}
  \label{fig:diff_priority_line}
\end{figure}

\begin{figure}[t]
  \centering
  \includegraphics[width=\linewidth]{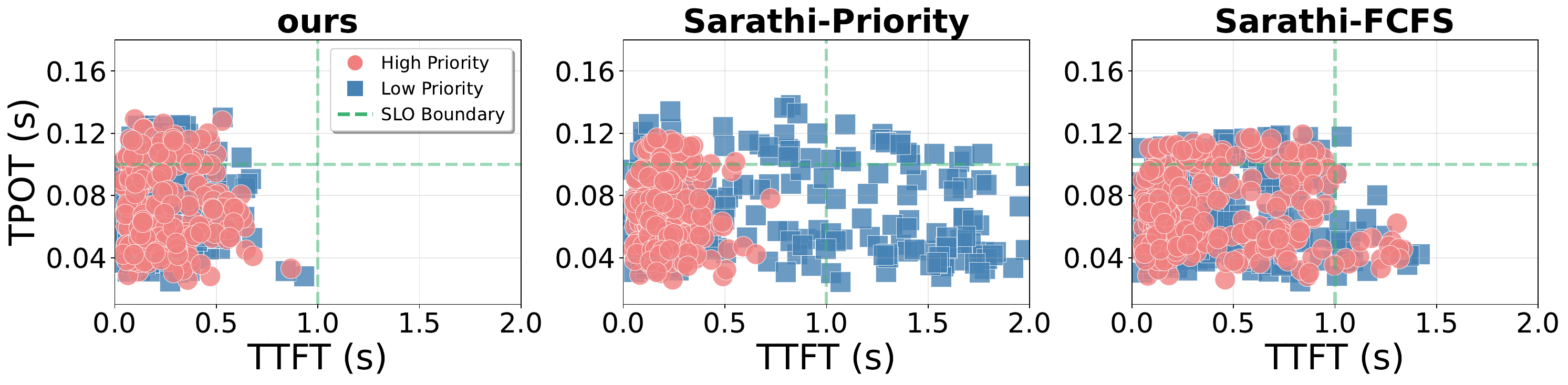}
  \caption{TTFT and TPOT distributions for different priority requests under various scheduling strategies.}
  \label{fig:diff_priority_scatter}
\end{figure}

\subsection{Performance of Different Priorities} \label{sec:diff_priority}

Figure~\ref{fig:diff_priority_line} compares performance across priority levels. \systemname{} preserves a desirable priority ordering (slightly higher TDG for high-priority requests) while keeping both priorities at high levels and consistently outperforming Sarathi-FCFS. In contrast, Sarathi-Priority over-favors high-priority requests, causing low-priority starvation and lower overall gain.
Figure~\ref{fig:diff_priority_scatter} further shows TTFT/TPOT distributions. \systemname{} maintains balanced latency across priorities with small disparity, while Sarathi-Priority yields much larger low-priority TTFT and many SLO violations. Although Sarathi's decode-prioritized design can slightly improve TPOT, it significantly increases TTFT timeout rates.

\begin{figure}[t]
  \centering
  \includegraphics[width=\linewidth]{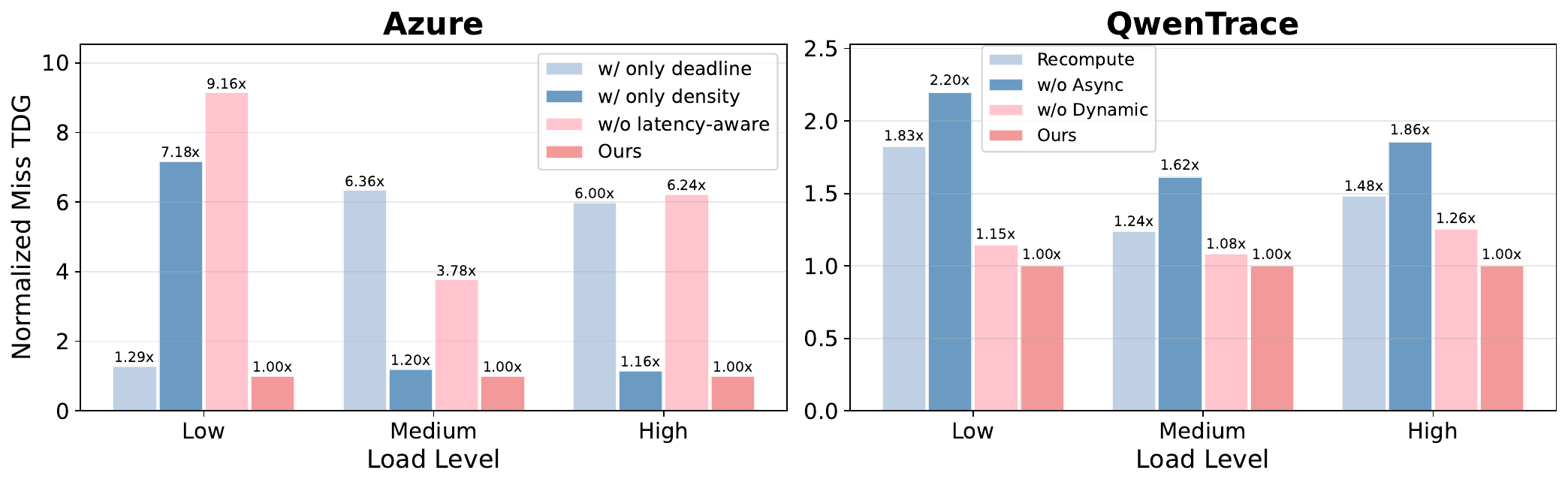}
  \caption{Left: Ablation study for \localsched{}. Right: Ablation study for efficient block management.}
  \label{fig:ablation}
\end{figure}

\subsection{Ablation Study} \label{ablation_study}

We conduct ablations to isolate each module in \systemname{}. In Figure~\ref{fig:ablation} (left), \textit{w/ only deadline} and \textit{w/ only density} remove Adaptive Urgency Partition and keep only one ordering strategy, while \textit{w/o latency-aware} disables the latency estimator for batch-capacity control. Removing any module degrades performance. \textit{w/ only deadline} is better at lower load, whereas \textit{w/ only density} becomes better at higher load, consistent with \textsection\ref{sec:static_scheduler}.
For efficient block management, we run a second ablation under a low memory-utilization threshold: \textit{w/o async} uses synchronous copying, \textit{w/o dynamic} always copies all host blocks, and \textit{Recompute} discards blocks upon eviction. Figure~\ref{fig:ablation} (right) confirms that each component contributes.

\begin{figure}[t]
  \centering
  \includegraphics[width=\linewidth]{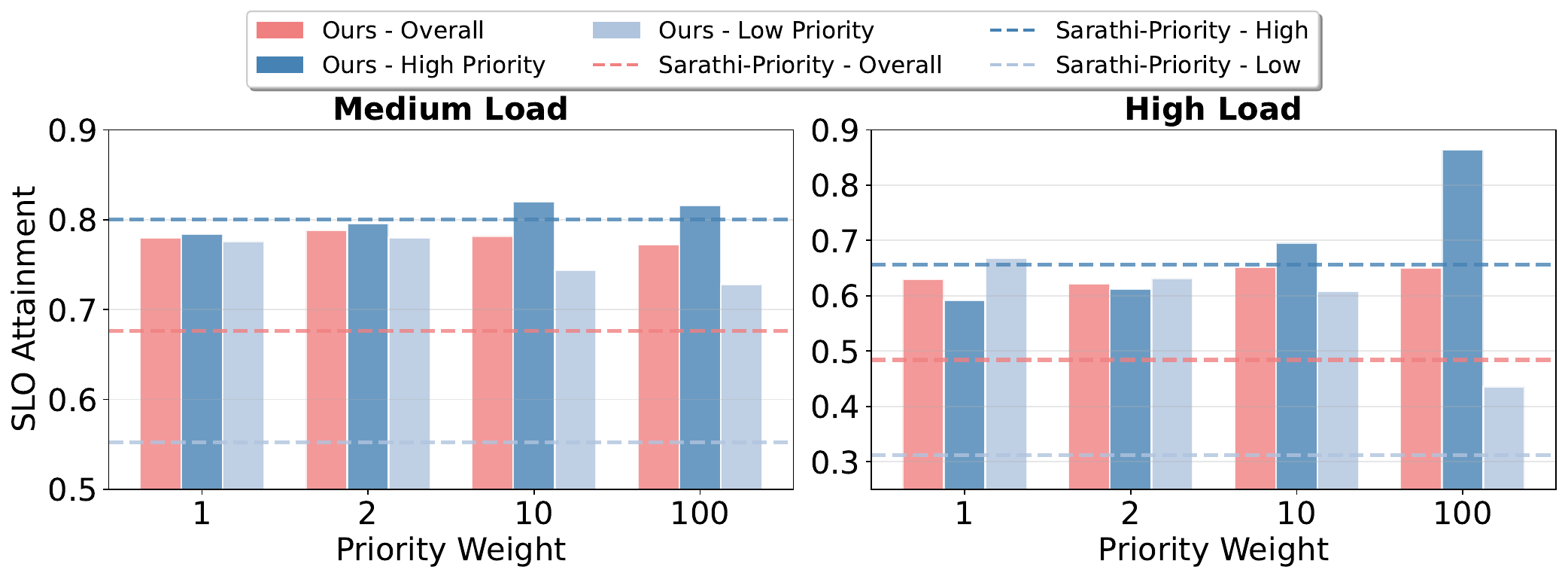}
  \caption{Effects of priority weight scaling on multi-priority request performance in \systemname{}.}
  \label{fig:priority_scale}
\end{figure}

\subsection{Priority Weight Scaling} \label{sec:priority_scale}

Figure~\ref{fig:priority_scale} reports SLO satisfaction under different priority weights and loads. As the priority weight increases, high-priority satisfaction rises while low-priority satisfaction declines, and overall satisfaction remains nearly stable, indicating that \systemname{} shifts service guarantees toward high-priority traffic without collapsing system-level performance. The gain for high-priority requests is larger under high load than under low-to-medium load, suggesting that priority-aware scheduling is most effective under contention. Compared with Sarathi-Priority, \systemname{} is initially slightly lower on high-priority satisfaction at small weights but surpasses it at larger weights, while consistently achieving better overall and low-priority satisfaction. As a result, \systemname{} demonstrates a better trade-off across priority levels rather than strict one-sided prioritization.




\begin{figure}[t]
  \centering
  \includegraphics[width=\linewidth]{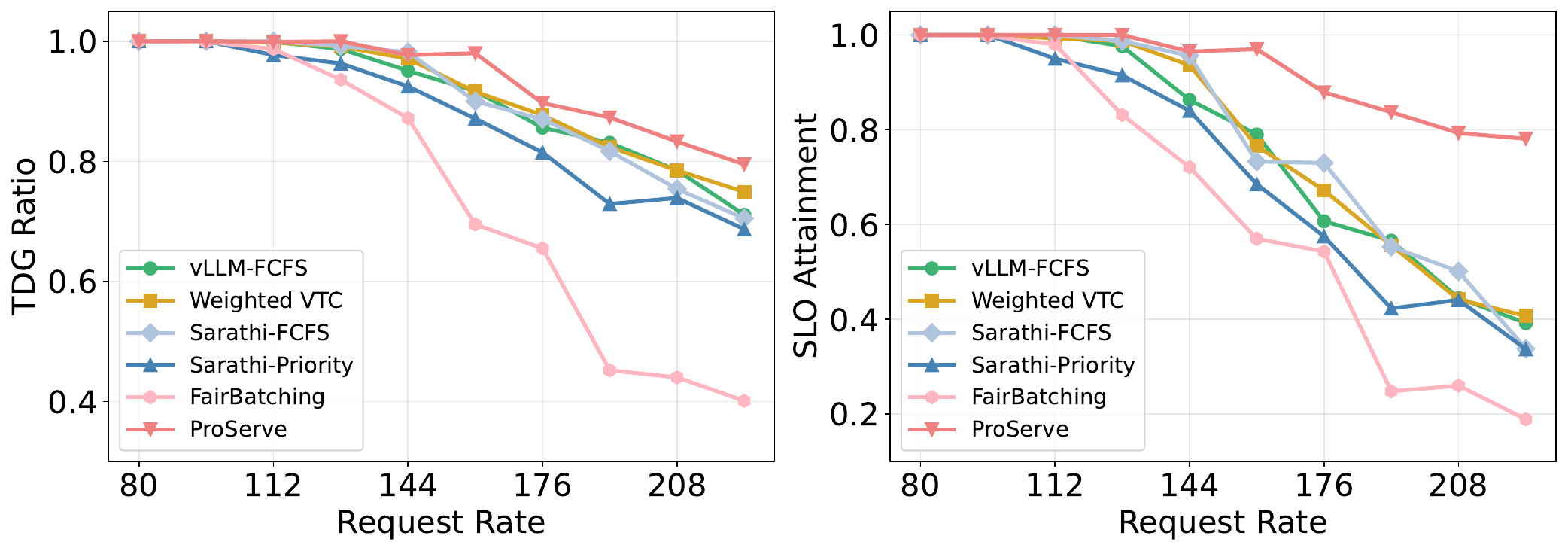}
  \caption{Performance of different methods on a large‑scale cluster using a proprietary industrial dataset.}
  \label{fig:large_cluster}
\end{figure}

\subsection{Large‑Scale Cluster Experiments}
\label{sec:cluster}

We conduct experiments using our proprietary industrial dataset (the distribution is shown in Figure~\ref{fig:multi_priority_load}). The priority weights for different priorities are assigned according to their actual business value in our production environment. We deploy 32 instances of the Qwen3‑32B model~\cite{qwen3} on 8 servers. 
The other baselines adopt the round-robin policy.
As shown in Figure~\ref{fig:large_cluster}, our method consistently outperforms all baseline methods even at this large scale and on the real‑world industrial workload.


\section{Related Work}

\noindent \textbf{LLM Serving.}
Prior work optimizes LLM serving from multiple angles, including kernel efficiency~\cite{kao2023flat, flashattention, flashattention2, zhang2024sageattention}, prefix caching~\cite{zheng2024sglang}, and KV cache management~\cite{kwon2023efficient, mooncake, li2024survey}. These system-level optimizations are orthogonal to us and can be integrated directly. On scheduling, Sarathi-Serve~\cite{sarathi-serve} uses chunked prefill and stall-free batching, while many recent methods focus on SLO-aware scheduling~\cite{huang2025slo, tang2025scorpio, sola, slos_serve, bin2025fineserve}.

\noindent \textbf{PD Disaggregation.}
DistServe~\cite{distserve} disaggregates prefill and decode to avoid cross-phase interference, and this architecture has been widely adopted~\cite{vllm, liu2025xllm}. Follow-up work improves it via parallelization~\cite{distserve}, KV management~\cite{mooncake}, and instance orchestration~\cite{pdagg_ornot, arrow}. Our method is compatible with PD disaggregation.

\noindent \textbf{Priority-related Request Scheduling.}
Recent schedulers often derive priority from request attributes such as length~\cite{fastserve}, SLO capability~\cite{liao2026laser,li2025adaserve}, or online/offline type~\cite{bros}. Some methods~\cite{fastserve, learning_to_rank, TetriInfer, prefillonly} prioritize short requests. 
Studies~\cite{tang2025scorpio, slos_serve, zhu2025polyserve} handle multi-SLO workloads by giving tighter-SLO requests higher priority, which covers only part of our setting. 
Works on online/offline co-location~\cite{wang2025echo, bros, sun2025hygen} typically prioritize online requests and treat offline traffic as best-effort. 
They generally disregard the latency
requirements of offline requests, making them unsuitable for
direct application in our scenario.
None explicitly model inherent priority differences among online clients.
Llumnix~\cite{sun2024llumnix} allocates more reserved memory space for high-priority requests. VTC~\cite{vtc} is a fairness-oriented scheduling algorithm. Its extension, Weighted VTC, introduces priority-specific weights to ensure that the ratio of processed tokens aligns with requests' priorities. However, memory reservation or static token quotas alone cannot explicitly guarantee latency for high-priority traffic.

\section{Conclusion}

In this paper, we first formalize the multi-priority scheduling problem as a service gain maximization task.
To address this, we propose \systemname{} consisting of: \localsched{}, which adaptively reorders requests according to load and priority; an efficient block management, which overlaps host-device transfers with computation; and \globalsched{}, which performs gain-oriented and capability-aware request dispatching.
Extensive experiments validate the effectiveness of \systemname{}.

\clearpage

\bibliographystyle{ACM-Reference-Format}
\bibliography{reference}

\clearpage

\appendix

\section{Other Details of GoRouting}
\label{gorouting_details}
The specific GoRouting algorithm is shown in Alg.~\ref{alg:request_dispatch_policy}.

\noindent \textbf{\textsc{EstimateGain}.}
GoRouting can be adapted to different local schedulers by changing the implementation of \textsc{EstimateGain}. We use the following unified form:
\begin{equation}
\begin{aligned}
\textsc{EstimateGain}(Q)
&=\sum_{r\in Q} w_r(1)\cdot\\
&\mathbb{I}\!\left[\textsc{EstimateExec}(Q,r)\le r.\text{remain}\right].
\end{aligned}
\end{equation}
The execution-time estimator \(\textsc{EstimateExec}(Q,r)\) is instantiated using the load-judgment function \(\phi\) described in \textsection\ref{sec:local_sched}. Specifically, in PD co-location, we uniformly use a conservative budget \(t_{\text{budget}}=\min_{r\in Q} TPOT_{\text{SLO}}^r\) to estimate execution time, together with the decode term \(\hat{t}_d\) added on top of the constant term \(t_c\). \(\textsc{EstimateExec}(Q,r)\) is computed based on the position of request \(r\) in the sorted queue \(Q\) (according to the specific local scheduler) by accumulating the execution costs of all requests ahead of \(r\).

\noindent \textbf{Estimated Decode Overhead in PD co-location.}
To extend execution-time estimation to PD co-location, we approximate the decode-side overhead using the decode estimator \(\tilde{T}_d(r)\) in \textsection\ref{sec:batch_estimator}. Specifically, we estimate a decode term:
\begin{equation}
\hat{t}_d(n_d) = a_d \cdot \hat{l}_{kv}^{\,d} + b_d \cdot n_d ,
\label{eq:appendix_decode_term}
\end{equation}
where \(n_d\) is the number of ongoing decode requests, and \(\hat{l}_{kv}^{\,d}\) is the estimated total KV cache token length generated by decode:
\begin{equation}
\hat{l}_{kv}^{\,d}=\left(M-b_f-\frac{L_{\text{pre}}}{s_{\text{blk}}}\right)\cdot s_{\text{blk}}.
\end{equation}
Here, \(M\) denotes the total number of KV blocks on an instance, \(b_f\) denotes the number of free blocks, \(s_{\text{blk}}\) is the number of tokens per block, and \(L_{\text{pre}}\) is the total input prefill-token length.

\section{Implementation Details}
\label{sec:appendix_implementation}
\systemname{} is implemented on top of the recently open-sourced high-performance LLM inference system xLLM~\cite{xllm}, which is written entirely in C++. xLLM adopts a service-engine decoupled architecture that aligns well with our design. We implement \globalsched{} at the service layer, and implement \localsched{} together with efficient hierarchical block management at the engine layer.
For PD disaggregated settings, we leverage xLLM's built-in KV-cache push mode, which asynchronously transfers KV caches from prefill instances to decode instances in a layer-wise manner~\cite{mooncake}. Consequently, GoRouting selects both the prefill and decode instances when each request arrives at the service layer.
For pipelined reloading, H2D transfers are launched on a dedicated copy stream independent of the main compute stream. We record a completion event for each layer chunk and synchronize on that event only before executing the corresponding chunk, enabling reloads of later chunks to overlap with computation of earlier chunks.
For asynchronous offloading, D2H transfers are issued on a separate copy stream and handled outside the critical forward path.

\section{Complexity Analysis of Service-Gain Maximization}
\label{sec:appendix_complexity}

\begin{theorem}[NP‑hardness of Gain Maximization]
\label{thm:nphard}
The unified multi‑priority request scheduling problem in ProServe, 
i.e., maximizing the total service gain $\max \sum_{r \in R} f_{\mathrm{TDG}}(r)$,
is NP‑hard.
\end{theorem}

\noindent \textbf{Proof.}
We prove NP-hardness by restriction.

\noindent \textbf{Step 1 (Restricted single-instance case).}
Consider the following \emph{single-instance} restricted special case of our problem:
(1) each request has only prefill computation (decode-side constraints are made non-binding),
(2) preemption is disallowed,
(3) at most one prefill request can be admitted at a time (batching is disabled by capacity constraints), and
(4) each request has a deterministic prefill processing time known in advance.

Under this restriction, each request \(r\) has arrival time \(a_r\), prefill processing time \(p_r\), TTFT slack \(s_r\), and priority weight \(w_{p(r)}\). The per-request gain is
\begin{equation}
f_{\mathrm{TDG}}(r)=w_{\mathrm{p}}\cdot w_{p(r)}\cdot \mathbb{I}[t_{\mathrm{prefill}}(r)\le s_r],
\end{equation}
where \(t_{\mathrm{prefill}}(r)\) is the elapsed time from arrival to prefill completion for request \(r\). Hence, the restricted objective is exactly the weighted number of on-time jobs on a single non-preemptive machine with release times and deadlines.

\noindent \textbf{Step 2 (Mapping from a classical scheduling problem).}
Now consider \(1\mid r_j,d_j\mid\sum_j w_j U_j\): each job \(j\) has release time \(r_j\), processing time \(p_j\), deadline \(d_j\), and weight \(w_j\); \(U_j=1\) iff job \(j\) is tardy (\(C_j>d_j\)), else \(U_j=0\). Minimizing \(\sum_j w_jU_j\) is equivalent to maximizing \(\sum_{j:C_j\le d_j} w_j\), i.e., the total weight of on-time jobs. This problem is NP-hard~\cite{lenstra1977complexity}.

Given any instance \(I\) of \(1\mid r_j,d_j\mid\sum_j w_j U_j\), we construct in polynomial time a restricted \systemname{} instance \(I'\) as follows:
\begin{itemize}
    \item For each job \(J_j\), create one request \(r_j\) with
    \begin{itemize}
        \item arrival time \(a_{r_j}=r_j\),
        \item prefill processing time \(p_{r_j}=p_j\),
        \item set the TTFT slack to \(d_j-r_j\) (jobs with \(d_j<r_j\) are trivially tardy),
        \item priority weight \(w_{p(r_j)}=w_j/w_{\mathrm{p}}\) (equivalently, set \(w_{\mathrm{p}}=1\) w.l.o.g.).
    \end{itemize}
    \item Enforce single-request admission (equivalently, disable batching via capacity constraints), so at most one request is processed at a time.
    \item Keep decode-side constraints non-binding (equivalently, prefill-only in this restricted case).
\end{itemize}

Because preemption is disabled and at most one request can run at a time, feasible schedules in \(I'\) are in one-to-one correspondence with feasible schedules in \(I\).

\noindent \textbf{Step 3 (Objective equivalence and hardness).}
For any corresponding schedule pair \((\pi_I,\pi_{I'})\), completion times are preserved under the mapping, i.e., \(C_j(\pi_I)=C_{r_j}(\pi_{I'})\) for all \(j\). Therefore,
\begin{equation}
\begin{aligned}
\sum_{r_j} f_{\mathrm{TDG}}(r_j;\pi_{I'})
&= \sum_j w_{\mathrm{p}}\,w_{p(r_j)}\,\mathbb{I}[t_{\mathrm{prefill}}(r_j;\pi_{I'})\le d_j-r_j] \\
&= \sum_j w_j\,\mathbb{I}[C_j(\pi_I)\le d_j] \\
&= \sum_j w_j\,(1-U_j(\pi_I)) \\
&= \sum_j w_j-\sum_j w_jU_j(\pi_I).
\end{aligned}
\end{equation}
Therefore, maximizing total TDG in the restricted \systemname{} instance is equivalent to minimizing \(\sum_j w_jU_j\) in \(1\mid r_j,d_j\mid\sum_j w_j U_j\).

Hence, this single-instance restricted \systemname{} problem is at least as hard as \(1\mid r_j,d_j\mid\sum_j w_j U_j\), which is NP-hard; therefore, the general TDG-maximization problem in \systemname{} is NP-hard. \(\square\)

\noindent \textbf{Implication.}
The above hardness already holds after removing many real-system complications (e.g., decode coupling and memory management). Therefore, exact online global optimization is intractable in general, motivating our load-adaptive heuristics.

\section{More Experimental Results}

\begin{figure}[t]
  \centering
  \includegraphics[width=\linewidth]{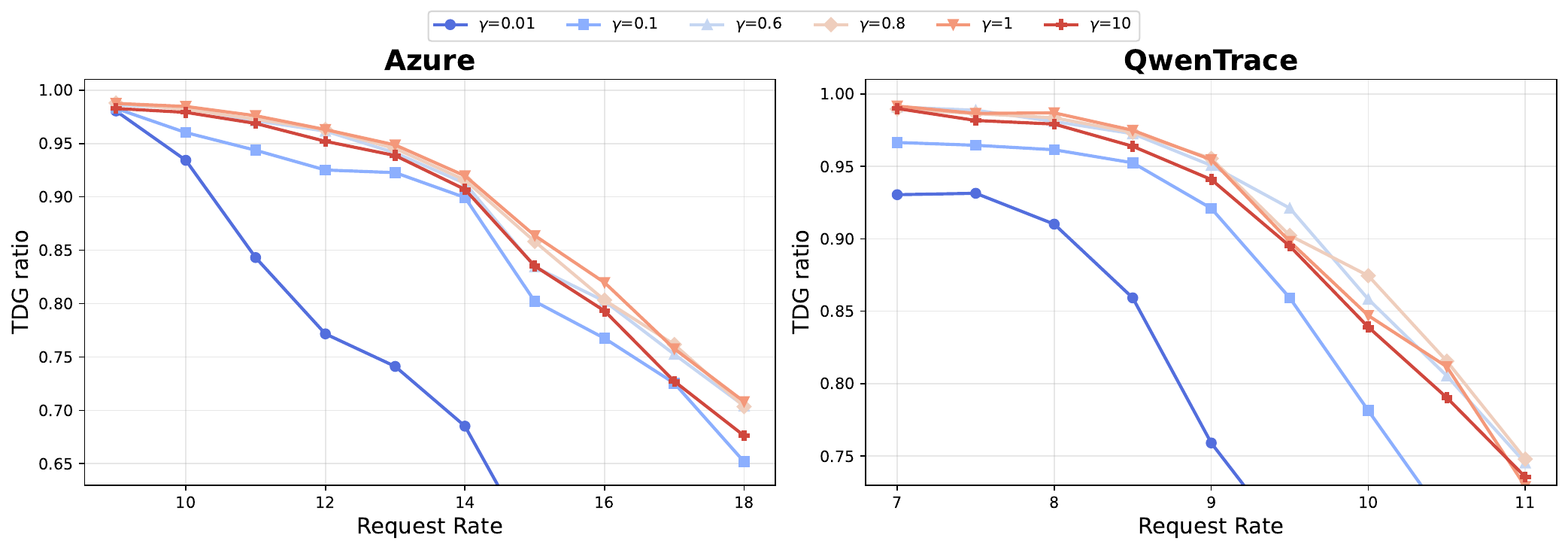}
  \caption{Effects of different aggressiveness coefficients in \localsched{}.}
  \label{fig:param_sensitivity}
\end{figure}

\subsection{Parameter Sensitivity Analysis} \label{sec:param_sensitivity}

Figure~\ref{fig:param_sensitivity} reports performance under different aggressiveness coefficients \(\gamma\) across multiple datasets.
First, as \(\gamma\) increases, overall performance generally follows an ``increase-then-decrease'' trend, with the best results achieved around \(0.8\) to \(1.0\) in our setup.
Second, except for \(\gamma = 0.01\), performance remains relatively stable across different \(\gamma\) values, indicating that \localsched{} is robust to moderate changes in \(\gamma\).
Finally, under higher loads, \(\gamma = 0.01\) degrades substantially.
This is consistent with our analysis in \textsection\ref{sec:static_scheduler}: when the scheduling strategy becomes overly EDF-like, cascading timeouts can occur and sharply reduce service gain.

\begin{figure}[t]
  \centering
  \includegraphics[width=\linewidth]{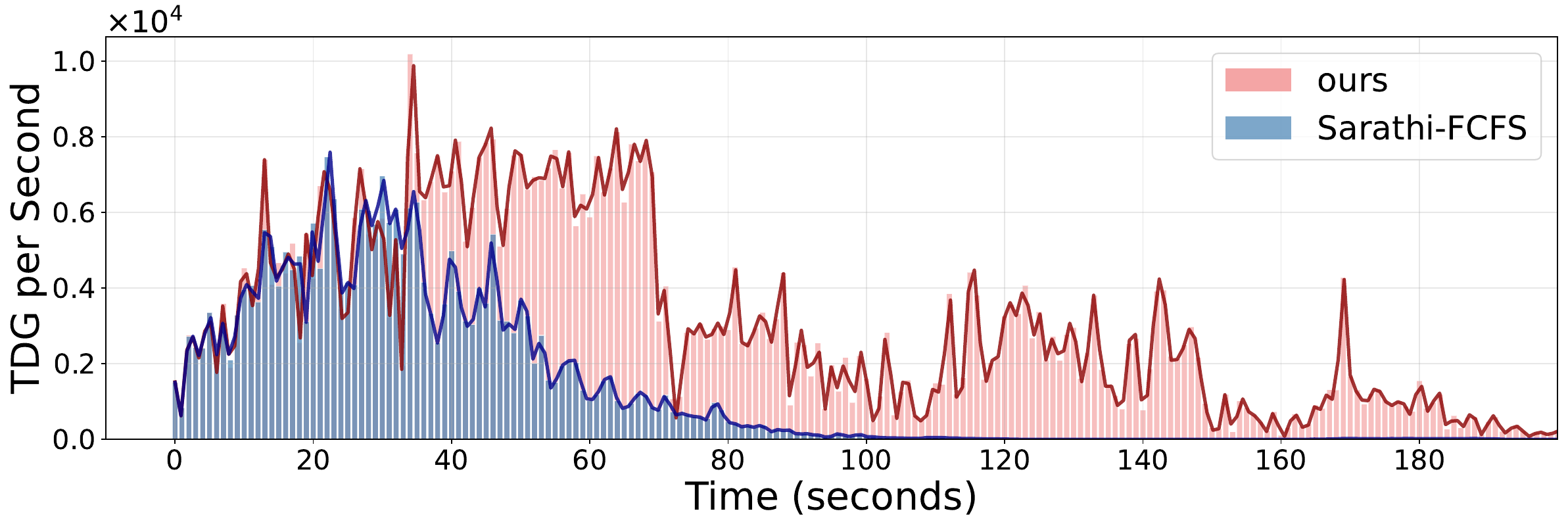}
  \Description{Timeline comparison of request servicing for different methods.}
  \vspace{-2.5em}
  \caption{The request servicing timelines of \systemname{} and baseline methods during a representative service session.}
  \label{fig:tdg_timeline}
\end{figure}

\begin{figure}[t]
  \centering
  \includegraphics[width=\linewidth]{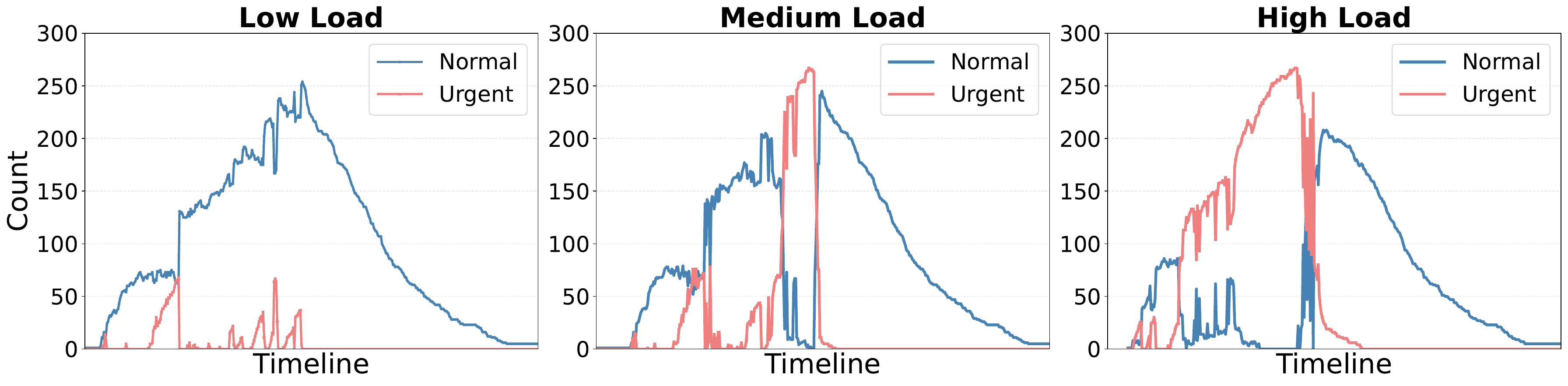}
  \vspace{-2.5em}
  \caption{The urgent and normal request distribution timelines of \systemname{} under different load (request rate).}
  \label{fig:urgent_count_timeline}
\end{figure}

\subsection{Timeline Analysis} \label{timeline_analysis}

Figure~\ref{fig:tdg_timeline} illustrates the timeline of TDG obtained per second by Sarathi-FCFS and \systemname{} under relatively high load on the Azure dataset. Sarathi-FCFS initially achieves TDG under low-load conditions; however, as the cumulative load increases, its FCFS-based scheduling can lead to widespread request timeouts, causing TDG to approach zero in subsequent service intervals. In contrast, \systemname{} can adaptively respond to load variations. It employs an approximately deadline-first strategy during low-load periods, achieving higher TDG than Sarathi-FCFS. More importantly, upon detecting high cumulative load conditions, \systemname{} dynamically switches to its high-load scheduling mode, prioritizing high-priority and relatively short requests to maintain sustained TDG acquisition throughout the service period.
Figure~\ref{fig:urgent_count_timeline} further illustrates the timeline of urgent and normal request counts partitioned by \localsched{} under different loads. It can be observed that our method adaptively adjusts the number of each type in response to load fluctuations. 

\subsection{Scheduler Overhead} \label{sec:schedule_overhead}
We further analyze the overhead of our scheduling algorithm, including both \localsched{} and \globalsched{}.
The overhead of \localsched{} is nearly identical to that of the traditional FCFS scheduler, accounting for only 0.17\% of the average batch forward execution time.
We also measure the fraction of \globalsched{} overhead in request TTFT, which increases from 0.04\% to 0.11\%.
This increase is still negligible.


\begin{table*}[!ht]
\centering
\resizebox{\textwidth}{!}{
\begin{tabular}{c|c|c|c|c|c|c|c}
\toprule
\textbf{Gain Function} & \textbf{Vanilla SLO} & \textbf{Weighted SLO} & \textbf{Tempo~\cite{zhang2025tempo}} & \textbf{Etalon~\cite{agrawal2024etalon}} & \textbf{Andes~\cite{liu2024andes}} & \textbf{TA-SLO} & \textbf{TDG(Ours)}\\
\midrule
Distinguishes Request Priority & \off & \on & \off & \off & \off & \on & \on \\
\midrule
Aware of Per-Token Latency & \off & \off & \on & \on & \on & \on & \on \\
\midrule
Distinguishes First/Decode Token Importance & \off & \off & \on & \off & \off & \on & \on \\
\midrule
Robust to Discard/Postpone Trick & \off/\off & \off/\off & \on/\off & \on/\off & \on/\on & \on/\off & \on/\on \\
\bottomrule
\end{tabular}
}
\caption{More detailed feature comparison of different per-request gain functions.}
\label{table:detailed_comparison_metric}
\end{table*}

\section{More Comparison of Gain Functions}
\label{sec:comparison}

Table~\ref{table:detailed_comparison_metric} presents a detailed comparison of representative gain formulations.
Tempo~\cite{zhang2025tempo} and Etalon~\cite{agrawal2024etalon} both adopt \emph{variable deadlines} by accumulating per-token deadlines.
Although this design can mitigate the early-drop trick, it still relies heavily on TBT-derived token deadlines and therefore remains vulnerable to the decode-postpone trick.
Our TA-SLO metric is inspired by this line of work and thus inherits the same limitation.
Andes~\cite{liu2024andes}, like our approach, uses a \emph{fixed deadline}. However, it does not account for request-level priority differentiation, nor does it explicitly distinguish first-token and decode-token utility (the former captures initial responsiveness, while the latter reflects output fluency).
In addition, Andes accumulates quality degradation from the user token-consumption perspective, where early violations are difficult to compensate later.
In contrast, we target \emph{multi-priority scheduling} through gain maximization, aggregating TDG from per-token gains.
Based on this objective, we co-design both engine-level and service-level schedulers: the load-adaptive \localsched{} and the gain-oriented \globalsched{}.
Thus, TDG serves not only as an evaluation metric but also as the driving force behind our multi-priority scheduling policies.



\end{document}